\documentstyle[preprint,aps]{revtex}
\begin{document}
\draft
\preprint{ANL-HEP-PR-94-74}
\title{Inclusive Prompt Photon Production in Hadronic\\
       Final States of $e^+e^-$ Annihilation }
\author{Edmond L. Berger$^1$, Xiaofeng Guo$^2$, and Jianwei Qiu$^2$}
\address{$^1$High Energy Physics Division,
             Argonne National Laboratory \\
             Argonne, Illinois 60439, USA \\
         $^2$Department of Physics and Astronomy,
             Iowa State University \\
             Ames, Iowa 50011, USA }
\date{July 27, 1995}
\maketitle

\begin{abstract}
We provide complete analytic expressions for the  inclusive prompt
photon production cross section in hadronic final states of
$e^+e^-$ annihilation reactions through one-loop order in quantum
chromodynamics perturbation theory.  Computed explicitly are direct
photon production through first order in the electromagnetic
strength $\alpha_{em}$ and the quark-to-photon and gluon-to-photon
fragmentation contributions through first order in the strong
coupling $\alpha_s$.  The full angular dependence of the cross
sections is displayed, separated into transverse
$(1+\cos^2\theta_\gamma)$ and longitudinal $(\sin ^2\theta_\gamma)$
components, where $\theta_\gamma$ specifies the direction of the
photon with respect to the $e^+e^-$ collision axis.  We discuss
extraction of fragmentation functions from $e^+e^-$ data.
\end{abstract}

\vspace{0.3in}


\section{Introduction}
\label{sec:intro}

Production of an energetic photon in association with hadrons probes the
short-distance dynamics of electron-positron, hadron-hadron, and
lepton-hadron reactions.  In addition to providing valuable tests of
perturbative quantum chromodynamics (pQCD), data from electron-positron
annihilation reactions permit
measurements of parton-to-photon fragmentation functions.

In QCD, the quark-photon collinear singularities that arise in each
order of perturbation theory, associated with
the hadronic component of the photon, are subtracted and absorbed into
quark-to-photon and gluon-to-photon fragmentation functions,
in accord with the factorization theorem \cite{AEMP}.
Fragmentation functions, $D\left( z, \mu^2\right)$, are inherently
nonperturbative quantities whose magnitude and dependence on
fractional momentum $z$ must be measured in experiments at a reference
fragmentation scale $\mu^2_0$.
The change of $D\left( z, \mu^2\right)$ with $\mu^2$
for large $\mu^2$ is specified
by perturbative QCD evolution equations \cite{WIT}.  In
$e^+e^- \rightarrow \gamma X$, the
fragmentation contributions play a significantly
greater role than they do in hadron-hadron collisions \cite{BXQ}.
In lowest-order, the quark-to-photon and anti-quark-to-photon
fragmentation processes dominate the inclusive reaction
$e^+e^- \rightarrow \gamma X$, whereas ``direct" processes,
such as $qg \rightarrow \gamma q$ and $q\bar{q} \rightarrow \gamma g$,
dominate in $pp \rightarrow \gamma X$ and $\bar{p}p \rightarrow \gamma X$
for $\gamma$'s that carry large values of transverse
momentum \cite{LO,NLO,BQ,DAT}.  The dominant role of fragmentation
contributions makes the inclusive process $e^+e^- \rightarrow \gamma X$
a potentially ideal source of information on
$D\left( z, \mu^2\right)$.

The most straightforward theoretical calculations in perturbative QCD
are those for the \underline{inclusive} yield of energetic photons,
$E_\gamma d\sigma/d^3p_\gamma$.  However, an important practical
limitation of high energy investigations is that photons are observed
and their cross sections are measured reliably only when the photons
are relatively isolated, separated to some extent in phase space from
accompanying hadrons.  Since fragmentation is a process in which
photons are part of quark, anti-quark, or gluon ``jets", it is evident
that photon isolation reduces the contribution from fragmentation terms.
In this paper, we deal with cross sections
for energetic \underline{inclusive} photons, and the extraction of
the photon fragmentation functions $D(z,\mu^2)$.
In another paper \cite{BXQ2}, we will present a systematic and analytic
treatment of cross sections for \underline{isolated} photons, and show
over what regions of $z$ the functions $D(z,\mu^2)$ may be determined
from data on isolated photon production in $e^+e^-\rightarrow \gamma
X$.  We will also point out the breakdown of factorization of the
cross section for isolated photons in a particular part of phase space in
$e^+e^- \rightarrow \gamma X$, and the implications of this breakdown
for calculations of isolated photon production in hadronic collisions.

Our calculations of the inclusive photon yields in
$e^+e^- \rightarrow \gamma X$
are carried out through one-loop order.  We compute explicitly direct
photon production through first order in the electromagnetic coupling
strength, $\alpha_{em}$, and the quark-to-photon and gluon-to-photon
fragmentation contributions through first order in the strong coupling
strength $\alpha_s$.  We display the full angular dependence of the
cross sections, separated into longitudinal $\sin^2 \theta_\gamma$ and
transverse components $\left( 1 + \cos^2 \theta_\gamma \right)$,
where $\theta_\gamma$ is the direction of the $\gamma$ with respect to
the $e^+e^-$ collision axis.  Our work goes beyond that of previous
authors \cite{GL,KRAM,LEP}.  For example, the full angular dependence
of the cross section was not derived before.
In one recent analysis \cite{GL}, the authors concentrate on
events having the topology of a photon plus
1 hadronic jet; they discuss the extraction of the quark to photon
fragmentation function from such data.  In that approach, final state
partons are treated as resolved, and the cancellation of infrared
singularities is not treated explicitly.  Practical aspects
of confronting theoretical calculations with data from LEP are
addressed in Ref.~\cite{LEP}.   All four groups at LEP have published
papers on prompt photon production \cite{LEP4}.  In this paper, we
advocate a different method of analysis of data from that used so far.

We begin in Section II with definitions of the factorized inclusive
photon cross sections, and, to establish notation, we derive explicit
expressions for the inclusive photon yields in lowest order
$\left( O\left( \alpha^o_{em}\right),\ O\left( \alpha^o_s\right) \right)$.
In Section III, we examine in turn the three 1st order contributions to the
inclusive photon yield: the $O\left( \alpha_{em}\right)$ process in
which a photon is radiated from a final quark or antiquark line,
$e^+e^- \rightarrow q\bar{q}\, \gamma$, and the $O\left( \alpha_s \right)$
processes in which $e^+e^- \rightarrow q\bar{q}\ g$, followed by
fragmentation of one of the three final-state partons into a photon.
The quark to photon collinear singularity in
$e^+e^- \rightarrow q\bar{q}\gamma$ is absorbed into the
quark-to-photon fragmentation function.  Our treatment of the
$O\left( \alpha_s\right)$ contributions necessarily includes a full
discussion of both real and virtual diagrams.  Dimensional regularization
is used to handle infrared and collinear singularities.  In Section IV, we
summarize our final expressions for the inclusive photon cross section
$E_\gamma d\sigma_{e^+e^- \rightarrow \gamma X}/d^3\ell$, with full
$\theta_\gamma$ dependence.  Numerical results and suggestions for
comparisons with $e^+e^-$ data at LEP, SLAC/SLC, TRISTAN, and CESR/CLEO
energies are also collected in Section IV.  An Appendix is included in
which we derive expressions for two- and three- particle phase space in
$n$ dimensions.

\section{Definitions, Notation, and Lowest Order Contribution}
\label{sec:def}

In this section we establish the notation to be used throughout the paper
and present our derivation of the lowest order
$O\left( \alpha^o_{em}\alpha^o_s\right)$ contribution to the inclusive
energetic photon yield in $e^+e^- \rightarrow \gamma X$.

\subsection{General Structure of the Cross Section and Kinematics}
\label{subsec:2a}

In $e^+e^-\rightarrow cX$, as sketched in Fig.~\ref{fig1},
the cross section for an $m$
parton final state is
\begin{equation}
d\sigma^{(m)} = {{1} \over {2s}}\Big| \overline{M}
_{e^+e^-\rightarrow {\underbrace{c+\cdots}_{m}}}
\Big|^2 dPS^{(m)}\cdot dz
D_{c\rightarrow \gamma} (z),
\label{l}
\end{equation}
with $c = \gamma, q, \bar{q}, g$ and $z = E_\gamma/E_c$.  For
inclusive photons, we integrate over all phase space, $dPS^{(m)}$,
except the momentum of parton ``$c$''.  For isolated photons, however,
the phase space, $dPS^{(m)}$, will have extra constraints due to the
definition of the isolated photon events.

For the scattering amplitude, $M_{e^+e^-\rightarrow c+\cdots}$, the
vertex between the intermediate vector boson and the initial/final
fermion pair is expressed
as\ i$e\gamma_\mu\, (v_f + a_f\, \gamma_5)$.  The absolute square
of the matrix element $|\overline{M}|^2$, averaged over initial spins and
summed over final spins and colors,
may be expressed in terms of leptonic and
hadronic tensors, $L_{\mu \nu}\ {\rm and}\ H^{\mu \nu}$, as
\begin{equation}
|\overline{M}|^2 = e^2C \left[ F^{PC} (q^2)\ L^{PC}_{\mu \nu} +
F^{PV} (q^2)\ L^{PV}_{\mu \nu}\right]  H^{\mu \nu};
\label{m}
\end{equation}
$e$ denotes the electric charge, and $C$ is the overall color factor.
Since the physical observable, the energetic photon $\gamma$, does not
distinguish between quarks and antiquarks, the parity violating $(PV)$
term does not contribute.  Equivalently, only the symmetric part of
$H^{\mu\nu}$ contributes.  Therefore,
\begin{equation}
|\overline{M}|^2 = e^2C F^{PC} (q^2)\ L^{PC}_{\mu \nu}\ H^{\mu \nu} \equiv
e^2C F^{PC}_q (q^2) \left( H_1 +H_2\right).
\label{n}
\end{equation}
\begin{equation}
H_1 = \left( -g_{\mu\nu}+{{q_\mu q_\nu}\over{q^2}} \right) H^{\mu \nu} =
- g_{\mu \nu} H^{\mu \nu}.
\label{o}
\end{equation}
\begin{equation}
H_2 = - {{k_\mu k_\nu} \over {q^2}} H^{\mu \nu}.
\label{p}
\end{equation}
The four-momenta $q^\mu$ and $k^\mu$ are defined in terms of the
four-momenta of the incident $e^+$ and $e^-\ \left( k^\mu_1 {\rm{and}}\
k^\mu_2 \right)$ as
\begin{equation}
q^\mu = k^\mu_1 + k^\mu_2,\ q^2 = \left( k_1 + k_2\right)^2 = s;
\label{q}
\end{equation}
and
\begin{equation}
k^\mu = k^\mu_1 - k^\mu_2,\ k^2 = \left( k_1 - k_2 \right)^2 = -s.
\label{r}
\end{equation}

The normalization factor $F^{PC}_q (q^2)$ is expressed in terms of
the vector $(v)$ and axial-vector $(a)$ couplings of the intermediate
$\gamma^*$ and $Z^o$ to the leptons and quarks.  At the $Z^o$ pole,
neglecting  $\gamma, Z^o$ interference, we find
\begin{equation}
{{2} \over {s}}\ F^{PC}_q (s) = \left(|v_e|^2 + |a_e|^2\right)
\left(|v_q|^2 + |a_q|^2\right)
{{1} \over {\left( s-M^2_Z\right)^2 + M^2_Z \Gamma^2_Z}}.
\label{s}
\end{equation}
At modest energies where only the $\gamma^*$ intermediate state is
relevant,
\begin{equation}
{{2} \over {s}}\ F^{PC}_q\ (s) = e^2_q\ {{1} \over {s^2}};
\label{t}
\end{equation}
$e_q$ is the fractional quark charge
$\left( e_u = 2/3; e_d = 1/3;\cdots\right)$.

In terms of functions $H_1$ and $H_2$, defined through Eq.~(\ref{n}),
we reexpress the cross section as
\begin{equation}
d\sigma^{(m)}= \sum_q \left[\frac{2}{s}F^{PC}_q(s)\right]\,
               e^2\, C\, \frac{1}{4}\left( H_1 +H_2\right)
               dPS^{(m)}\, dz\, D(z)\ .
\label{dsigma2}
\end{equation}
In the sections to follow, we calculate the functions $H_1$ and $H_2$
explicitly for the lowest order and the first-order contributions to
$e^+e^- \rightarrow \gamma X$.  Function $H_1$ provides the cross section
integrated over all production angles, $\theta_\gamma$, of the $\gamma$.
Function $H_2$ specifies the angular dependence or, equivalently, the
transverse momentum distribution of the $\gamma$ with respect to the
$e^+e^-$ collision axis.

\subsection{Factorized Cross Section}
\label{subsec:2b}

We are interested in the inclusive cross section for production of
photons in association with hadrons,  $E_\gamma d\sigma^{incl}_{e^+e^-
\rightarrow \gamma X}/d^3\ell$, where $E_\gamma$ is the energy of the
photon, and $\ell$ is the momentum of the photon in the  $e^+e^-$
center-of-mass system.  According to the pQCD factorization
theorem \cite{AEMP}, we may express the cross section as
\begin{equation}
E_\gamma{{d\sigma^{incl}_{e^+e^-\rightarrow \gamma X}} \over
{d^3\ell}} \equiv \sum_c E_c{{d\hat{\sigma}^{incl}_{e^+e^-\rightarrow cX}}
\over {d^3p_c}} \otimes D_{c \rightarrow \gamma} (z).
\label{a}
\end{equation}
The intermediate partons are $c = \gamma, g, q$, and $\bar{q}$.  The
hard-scattering cross section $E_c d\hat{\sigma}^{incl}_{e^+e^-
\rightarrow cX}/d^3p_c$ contains no infrared or collinear divergences.
The fractional momentum $z$ is defined as $z = E_\gamma/E_c$; all
intermediate partons $c$ are assumed to be massless.  The fragmentation
functions $D_{c\rightarrow \gamma} (z)$ represent all long-distance
physics associated with the hadronic component of the photon.  They
are inherently non-perturbative quantities that must be
measured experimentally.  Models and phenomenological
parametrizations \cite{JFO}
for $D(z)$ have been published.  In lowest-order,
$D_{\gamma \rightarrow \gamma} (z) = \delta (1-z)$.
The convolution expressed in Eq.~(\ref{a}) is sketched in Fig.~\ref{fig2}.
The symbol
$\otimes$ in Eq.~(\ref{a}) is defined explicitly as follows:
\begin{equation}
E_c{ {d\hat{\sigma}^{incl}_{e^+e^- \rightarrow cX}} \over
{d^3p_c}} \otimes D_{c \rightarrow \gamma} (z) \equiv \int^1_{z_{\min}}
{{dz} \over {z^2}}
\left[ E_c {{d\hat{\sigma}^{incl}_{e^+e^- \rightarrow cX}
\left( E_c = {{E_\gamma} \over {z}}\right)} \over
{d^3p_c}}\right]
D_{c\rightarrow \gamma}(z).
\label{b}
\end{equation}
Since $z_{\min}$ occurs when $p_c$ has its maximum value,
$p_c^{\max} = \sqrt{s}/2$, the lower limit of integration
$z_{\min} = x_\gamma = 2E_\gamma/  \sqrt{s}$; $\sqrt{s}$ is the center
of mass energy of the $e^+e^-$ annihilation.

\subsection{Derivation of the Lowest Order Contribution}
\label{subsec:2c}

In this section we present an explicit derivation of the lowest order
contribution to the inclusive photon yield in $e^+e^- \rightarrow \gamma$,
sketched in Fig.~\ref{fig3}.  The differential inclusive cross section
$d\sigma_{e^+e^- \rightarrow \gamma X}$
is expressed as a product of the lowest
order partonic cross section
$d\hat{\sigma}^{(o)}_{e^+e^- \rightarrow q\bar{q}}$
and the $q \rightarrow \gamma$ fragmentation function,
$D_{q \rightarrow \gamma} (z)$.
\begin{equation}
d\sigma_{e^+e^- \rightarrow \gamma X} = \sum_q d\hat{\sigma}^{(o)}_
{e^+e^- \rightarrow q(p_q)\bar{q}}\ dz\ D_{p_q \rightarrow \gamma} (z) +
(q \rightarrow \bar{q}).
\label{u}
\end{equation}
In Eq.~(\ref{u}), $p_q$ is the four-vector momentum of the quark $q$, and
$z = \ell/p_q$.  The partonic cross section is written, in turn, in terms of
the invariant matrix element and differential phase space factor.
\begin{eqnarray}
d\hat{\sigma}^{(o)}_{e^+e^- \rightarrow p_q p_{\bar{q}}}
&=& {{1} \over {2s}}
   \Big| \overline{M}_{e^+e^- \rightarrow p_q p_{\bar{q}}} \Big|^2\
   dPS^{(2)} \nonumber \\
&=& \left[\frac{2}{s}F^{PC}_q(s)\right]\,
   e^2\, N_c\, \frac{1}{4}\left( H_1 +H_2\right)\, dPS^{(2)}\ ,
\label{v}
\end{eqnarray}
where $N_c = 3$ is the number of colors carried by the quarks, and
Eq.~(\ref{n}) was used.

The symmetric part of the hadronic tensor $H^{\mu\nu}$, used to define
functions $H_1$ and $H_2$, is particularly simple:
\begin{equation}
H^{\mu\nu} = 4 \left(e\mu^{\epsilon}\right)^2
             \left[ p_{q}^{\mu} p_{\bar{q}}^{\nu}
                  + p_{\bar{q}}^{\mu} p_{q}^{\nu}
                  - g^{\mu\nu} p_q \cdot p_{\bar{q}} \right].
\label{w}
\end{equation}
The factor $\mu^{\epsilon}$ in Eq.~(\ref{w}) accommodates the fact that
we are working in $n$ dimensions.  The dimensional scale $\mu$ will be
specified further below.
The functions $H_1$ and $H_2$, defined in
Section~\ref{subsec:2a}, become
\begin{mathletters}
\label{x}
\begin{eqnarray}
H_1 &=&  4\left(e\mu^{\epsilon}\right)^2\,
        s\, (1-\epsilon)\ ;  \label{x1} \\
H_2 &=& -2\left(e\mu^{\epsilon}\right)^2\,
        s\, (1 - \cos^2 \theta)\ . \label{x2}
\end{eqnarray}
\end{mathletters}
In Eq.~(\ref{w}), $\epsilon$ is defined through the number of space
dimensions $n = 4-2\epsilon$, with $\epsilon \rightarrow 0$ at the
end of the calculation. In the center of mass frame of the collision,
$\theta$ is the angle of
$\vec{p}_q$ with respect to the direction defined by the incident $e^+$.
Combining $H_1$ and $H_2$, we obtain
\begin{equation}
\frac{1}{4}\left(H_1+H_2\right)=
\frac{1}{2}\, \left(e\mu^\epsilon\right)^2\, s\,
\left[ (1 + \cos^2 \theta) - 2\epsilon \right]\ .
\label{y}
\end{equation}
\noindent
Combining Eqs.~(\ref{y}) and the expression for
two-particle phase space in $n$-dimensions, Eq.~(\ref{A.4})
of the Appendix,
we find that the lowest order partonic cross
section, Eq.~(\ref{v}), is
\begin{equation}
E_q{{ d\hat{\sigma}^{(0)}_{e^+e^-\rightarrow qX}} \over {d^3 p_q}} =
\left[{{2} \over {s}} F^{PC}_q (s)\right] \alpha^2_{em} N_c
\left( {{4 \pi \mu^2} \over {(s/4) \sin^2\theta}}\right)^\epsilon
{{1} \over {\Gamma(1-\epsilon)}}
\left[ \left( 1 + \cos^2\theta \right) -2\epsilon \right]
{{\delta(x_q-1)} \over {x_q}},
\label{dd}
\end{equation}
with $x_q = 2E_q/\sqrt{s}$.
At this order, the cross section is manifestly finite in the limit
$\epsilon \rightarrow 0$, and we may set $\epsilon = 0$ directly in
Eq.~(\ref{dd}).  Nevertheless, Eq.~(\ref{dd}) expressed in $n$ dimensions
is valuable for later comparison with the higher order cross section.

Noting that
$\ell = zp_q$ implies $d^3p_q/E_q = \left( 1/z^2\right) d^3\ell/
E_\gamma$, we obtain the lowest order inclusive cross section
\begin{eqnarray}
E_\gamma {{d\sigma^{incl}_{e^+e^-\rightarrow \gamma X}} \over d^3 \ell}
&=& 2\sum_q\ \int^1\ {{dz} \over {z^2}}
\left[ E_q{{ d\hat{\sigma}^{(0)}_{e^+e^-\rightarrow q X}}
\over {d^3p_q}} \left( x_q = {{x_\gamma} \over {z}}\right)\right]
D_{q\rightarrow \gamma} (z, \mu_F)\nonumber \\
&=& 2\sum_q \left[ {{2} \over {s}} F^{PC}_q (s)\right] \alpha^2_{em}
N_c (1 + \cos^2\theta_\gamma) {{1} \over {x_\gamma}}
D_{q \rightarrow \gamma} (x_\gamma, \mu_F).
\label{ee}
\end{eqnarray}
The angles $\theta_\gamma$ and $\theta$ are identical since we take all
products of the fragmentation to be collinear.  The overall factor of 2 in
Eq.~(\ref{ee}) accounts for the $\bar{q}$ contribution.  In Eq.~(\ref{ee})
we have introduced a fragmentation scale $\mu_F$ in the specification of
the fragmentation function.

\section{First Order Contributions}
\label{sec:first}

There are three distinct contributions to $e^+e^-\rightarrow \gamma X$
in first order perturbation theory:
\begin{mathletters}
\label{ii}
\begin{eqnarray}
e^+e^- \rightarrow \gamma, &\hspace{0.5in}& O(\alpha_{em})
\label{ii1} \\
e^+e^- \rightarrow q \
(\mbox{or}\ \bar{q}) \rightarrow \gamma, &\hspace{0.5in}& O(\alpha_s)
\label{ii2} \\
& \rm{and} & \nonumber \\
e^+e^- \rightarrow g \rightarrow \gamma. &\hspace{0.5in}& O(\alpha_s)
\label{ii3}
\end{eqnarray}
\end{mathletters}
In Eqs. (\ref{ii2}) and (\ref{ii3}),
we have in mind contributions from quark and gluon
fragmentation to photons in the three-parton final state process
$e^+e^- \rightarrow q\bar{q}g$.  The first contribution, Eq.~(\ref{ii1}),
arises
from $e^+e^- \rightarrow q\bar{q}\gamma$ where the $\gamma$ is
\underline{not} collinear with either $\bar{q}$ or $q$.

In this section we derive and present the explicit contributions to the
inclusive yield
$E_\gamma d\sigma^{incl}_{e^+e^- \rightarrow\gamma X}/d^3\ell$
from each of the three processes in Eq.~(20).  Following the pQCD
factorization theorem, and Eq.~(\ref{a}), we must calculate the
short-distance hard-scattering cross sections,
$E_c d\hat{\sigma}^{incl}_{e^+e^- \rightarrow cX}/d^3p_c$
for $c=\gamma,g,q$ and $\bar{q}$.

The Feynman graphs for $e^+e^- \rightarrow\gamma q\bar{q}$ are sketched in
Fig.~\ref{fig4}.  Owing to the quark-photon
collinear divergence, the cross section
associated with these graphs is formally divergent.  We denote this
divergent first order cross-section
$\sigma^{(1)}_{e^+e^- \rightarrow\gamma X}$,
a short-hand notation for $Ed\sigma/d^3\ell$.
To derive the corresponding short-distance hard-scattering cross
section, $\hat{\sigma}^{(1)}_{e^+e^- \rightarrow\gamma X}$,
we apply the factorized form, Eq.~(\ref{a}), perturbatively,
\begin{eqnarray}
\sigma^{(1)}_{e^+e^- \rightarrow\gamma X} &=& \hat{\sigma}^{(1)}_
{e^+e^- \rightarrow\gamma X} \otimes D^{(0)}_{\gamma\rightarrow\gamma} (z)
\nonumber \\
&+& \hat{\sigma}^{(0)}_{e^+e^- \rightarrow q X} \otimes
D^{(1)}_{q\rightarrow \gamma} (z)\nonumber \\
&+& (q \rightarrow \bar{q}).
\label{jj}
\end{eqnarray}
The convolution represented by $\otimes$ is defined in Eq.~(\ref{b}).  The
superscripts (0) and (1) on the hard-scattering cross sections
$\hat{\sigma}$ and fragmentation functions $D$ refer to lowest-order
and first order, respectively.  The collinear divergence resides
in the first order fragmentation function
$D^{(1)}_{q\rightarrow \gamma}(z)$.
The hard-scattering cross sections $\hat{\sigma}^{(1)}$ and
$\hat{\sigma}^{(0)}$ are finite.
The expression for $\hat{\sigma}^{(0)}$ was derived in
Section~\ref{subsec:2c}.  In
Section~A we present our derivation of
$\hat{\sigma}^{(1)}_{e^+e^- \rightarrow\gamma X}$:
\begin{equation}
\hat{\sigma}^{(1)}_{e^+e^- \rightarrow \gamma X} =
\sigma^{(1)}_{e^+e^- \rightarrow \gamma X} -
\hat{\sigma}^{(0)}_{e^+e^- \rightarrow q X} \otimes
D^{(1)}_{q\rightarrow \gamma} (z) - (q \rightarrow \bar{q}).
\label{kk}
\end{equation}

The two Feynman graphs that provide the cross section for
$e^+e^- \rightarrow g \rightarrow \gamma$ in\linebreak
 $O(\alpha_s)$ are shown in
Fig.~\ref{fig5}.  In this
case, the final gluon is effectively ``observed'' through the fragmentation
$g\rightarrow \gamma$; there are no virtual gluon exchange diagrams.  The
finite hard-scattering cross section
$\hat{\sigma}^{(1)}_{e^+e^- \rightarrow g X}$ is derived from
the difference
\begin{equation}
\hat{\sigma}^{(1)}_{e^+e^- \rightarrow g X} =
\sigma^{(1)}_{e^+e^- \rightarrow g X} -
\sum_{q^\prime =q}^{\bar{q}}
\hat{\sigma}^{(0)}_{e^+e^- \rightarrow q^\prime X} \otimes
D^{(1)}_{q^\prime \rightarrow g}.
\label{ll}
\end{equation}
In Eq.~(\ref{ll}), the divergent cross section
$\sigma^{(1)}_{e^+e^- \rightarrow g X}$
is evaluated from the Feynman graphs shown in Fig.~\ref{fig5},
and the quark-to-gluon
collinear divergences are embedded in the first-order fragmentation function
$D^{(1)}_{q^\prime \rightarrow g}$.  Our derivation of
$\hat{\sigma}^{(1)}_{e^+e^- \rightarrow g X}$ is presented in Section~B.

The Feynman graphs in $O(\alpha_s)$ that contribute to
$e^+e^- \rightarrow q \rightarrow \gamma$ (Eq.~(\ref{ii2}))\linebreak
are sketched in Fig.~\ref{fig6}.  Only a final state
photon from quark fragmentation is observed.  The complete $O(\alpha_s)$
result includes both real gluon emission and virtual gluon exchange graphs,
as shown in Fig.~\ref{fig6}.
Although infrared divergences associated with soft gluons
cancel between the real and virtual graphs, the
cross section $\sigma^{(1)}_{e^+e^- \rightarrow q X}$
obtained from the Feynman graphs is still divergent due to collinear
singularities when the real gluon is emitted along the direction of its
parent quark or antiquark.  To obtain the corresponding hard-scattering
cross section $\hat{\sigma}^{(1)}_{e^+e^- \rightarrow q X}$, we apply
the factorized form, Eq.~(\ref{a}), perturbatively, to the production of
a quark instead of the photon,
\begin{equation}
\sigma^{(1)}_{e^+e^- \rightarrow q X} = \sum^{\bar{q}}_{q^\prime=q}
\hat{\sigma}^{(0)}_{e^+e^- \rightarrow q^\prime} \otimes
D^{(1)}_{q^\prime \rightarrow q} + \hat{\sigma}^{(1)}
_{e^+e^- \rightarrow q X} \otimes D^{(0)}_{q\rightarrow q},
\label{mm}
\end{equation}
with the collinear $q^\prime \rightarrow q$ singularities in
$O(\alpha_s)$ included in $D^{(1)}_{q^\prime \rightarrow q}$.  Note that
$D^{(0)}_{q\rightarrow q} (z) = \delta(1-z)$.
Correspondingly, the finite hard-scattering cross section
$\hat{\sigma}^{(1)}_{e^+e^- \rightarrow q X}$ is
\begin{equation}
\hat{\sigma}^{(1)}_{e^+e^- \rightarrow q X} =
\sigma^{(1)}_{e^+e^- \rightarrow q X} -
\hat{\sigma}^{(0)}_{e^+e^- \rightarrow q} \otimes
D^{(1)}_{q\rightarrow q}.
\label{nn}
\end{equation}
In Section C we present a detailed derivation of
$\hat{\sigma}^{(1)}_{e^+e^- \rightarrow q X}$.

Before turning to our explicit derivations, we conclude this discussion
with a presentation of the factorized formula for
the two-loop short-distance hard-scattering cross sections.
The two-loop direct contribution to $e^+e^-\rightarrow \gamma X$
is of $O(\alpha_{em}\alpha_s)$ and can be derived as follows.
We first apply the factorized form, Eq.~(\ref{a}),
perturbatively, at two-loop level and sum over $c=\gamma, g, q$ and
$\bar{q}$,
\begin{eqnarray}
\sigma^{(2)}_{e^+e^- \rightarrow\gamma X}
&=& \hat{\sigma}^{(2)}_{e^+e^- \rightarrow\gamma X}
   \otimes D^{(0)}_{\gamma\rightarrow\gamma} (z)
 + \hat{\sigma}^{(1)}_{e^+e^- \rightarrow\gamma X}
   \otimes D^{(1)}_{\gamma\rightarrow\gamma} (z)\nonumber \\
&+& \hat{\sigma}^{(1)}_{e^+e^- \rightarrow g X}
   \otimes D^{(1)}_{g\rightarrow \gamma} (z)
 + \hat{\sigma}^{(0)}_{e^+e^- \rightarrow g X}
   \otimes D^{(2)}_{g\rightarrow \gamma} (z)\nonumber \\
&+& \hat{\sigma}^{(1)}_{e^+e^- \rightarrow q X}
   \otimes D^{(1)}_{q\rightarrow \gamma} (z)
 + \hat{\sigma}^{(0)}_{e^+e^- \rightarrow q X}
   \otimes D^{(2)}_{q\rightarrow \gamma} (z)\nonumber \\
&+& (q \rightarrow \bar{q}).
\label{eeg2}
\end{eqnarray}
All first-order contributions, $\hat{\sigma}^{(1)}$,
in Eq.~(\ref{eeg2})
are given in Eqs.~(\ref{kk}), (\ref{ll}) and (\ref{nn}),
and they are calculated in this paper.  Since the first order
fragmentation functions $D^{(1)}_{\gamma\rightarrow \gamma}(z)$
and $D^{(1)}_{g\rightarrow \gamma}(z)$ vanish, and the zeroth order
hard-scattering cross section
$\hat{\sigma}^{(0)}_{e^+e^- \rightarrow g X}$ vanishes, we
derive the two-loop hard-scattering cross sections
$\hat{\sigma}^{(2)}_{e^+e^- \rightarrow\gamma X}$ as:
\begin{eqnarray}
\hat{\sigma}^{(2)}_{e^+e^- \rightarrow \gamma X}
&=& \sigma^{(2)}_{e^+e^- \rightarrow \gamma X} \nonumber \\
&-& \hat{\sigma}^{(1)}_{e^+e^- \rightarrow q X}
   \otimes D^{(1)}_{q\rightarrow \gamma} (z)
 - \hat{\sigma}^{(0)}_{e^+e^- \rightarrow q X}
   \otimes D^{(2)}_{q\rightarrow \gamma} (z) \nonumber \\
&-& (q \rightarrow \bar{q}).
\label{hs2g}
\end{eqnarray}
To complete the calculation of
$\hat{\sigma}^{(2)}_{e^+e^- \rightarrow\gamma X}$, it is necessary
to calculate the two-loop parton-level cross section
$\sigma^{(2)}_{e^+e^- \rightarrow\gamma X}$ and the two-loop
quark-to-photon fragmentation function
$D^{(2)}_{q\rightarrow \gamma} (z)$ in $n$-dimensions (implicitly, we use
dimensional regularization), in addition to all the zeroth and first
order contributions calculated in this paper.
The two-loop parton level cross section
$\sigma^{(2)}_{e^+e^- \rightarrow\gamma X}$ is formally divergent.
As is true of the calculation of
$\hat{\sigma}^{(1)}_{e^+e^- \rightarrow q X}$ in
Section~\ref{subsec:3c},
all infrared divergences associated with soft gluons
cancel among the real emission and virtual exchange diagrams.
All collinear divergences that
appear when final-state quarks and/or gluons are parallel
to the observed photon are cancelled by the subtraction terms given
in Eq.~(\ref{hs2g}).  Consequently, the two-loop hard-scattering
cross section $\hat{\sigma}^{(2)}_{e^+e^- \rightarrow \gamma X}$
is finite if the pQCD factorization theorem holds.

As shown in Section IV, the leading order short-distance
direct production contribution
$\hat{\sigma}^{(1)}_{e^+e^- \rightarrow \gamma X}$
is much smaller than the leading order fragmentation
contribution\linebreak
$\hat{\sigma}^{(0)}_{e^+e^- \rightarrow qX}\otimes
 D^{(1)}_{q\rightarrow\gamma}(z) + (q\leftrightarrow\bar{q})$.
We expect that the next-to-leading order direct contribution
$\hat{\sigma}^{(2)}_{e^+e^- \rightarrow \gamma X}$ will be much smaller
than the next-to-leading order fragmentation contributions
$\hat{\sigma}^{(1)}_{e^+e^- \rightarrow cX}\otimes
 D^{(1)}_{c\rightarrow\gamma}(z)$ with $c=g, q$ and $\bar{q}$,
which are completely derived in this paper.  We will not
calculate the two-loop contributions in this paper because we believe
their contributions to the overall cross section are much too small
in comparison with those presented here.

\subsection{Derivation of
$\hat{\sigma}^{(1)}_{e^+e^- \rightarrow \gamma X}$}
\label{subsec:3a}

In this section we present an explicit derivation of the finite
hard-scattering
cross section $\hat{\sigma}^{(1)}_{e^+e^- \rightarrow \gamma X}$ in
$O(\alpha_{em})$.
We begin by computing the functions $H_1$ and $H_2$, defined in
Section~\ref{subsec:2a},
Eqs.~(\ref{o}) and (\ref{p}).  These will then be integrated over phase
space to yield the cross section
$E_\gamma d\sigma^{(1)}_{e^+e^- \rightarrow \gamma X}/d^3\ell$:
\begin{equation}
d\sigma^{(1)}_{e^+e^-\rightarrow \gamma X} =
\sum_{q} \left[ {{2} \over {s}}\, F^{PC}_q(s)\right]
e^2\, N_c\, \frac{1}{4}(H_1 + H_2) dPS^{(3)},
\label{oo}
\end{equation}
where three-particle phase space in $n$-dimensions is given in
Eq.~(\ref{A.22a}) of the Appendix.

Sketched in Fig.~\ref{fig7} is the hadronic tensor
$H_{\mu\nu}$ obtained from the two
diagrams of Fig.~\ref{fig4}.
Performing traces to sum over final spins, we may write
the four contributions as
\begin{eqnarray}
H^{(a)}_{\mu\nu} &=& 2(1-\epsilon) Tr \left[ \gamma_\mu \gamma\cdot\ell
\gamma_\nu \gamma\cdot p_2\right] {1 \over {2p_1\cdot\ell}};\nonumber \\
H^{(b)}_{\mu\nu} &=& 2(1-\epsilon) Tr \left[ \gamma_\mu \gamma\cdot p_1
\gamma_\nu \gamma\cdot\ell\right] {1 \over {2p_2 \cdot\ell}};\nonumber \\
H^{(c)}_{\mu\nu} &=& -2Tr \left[ \gamma_\mu \gamma\cdot p_1 \gamma\cdot
p_2 \gamma_\nu \gamma\cdot (p_1 + \ell) \gamma\cdot (p_2 + \ell)\right]
{1 \over{2p_1 \cdot \ell}}\,  {1 \over{2p_2 \cdot {\ell}}}\nonumber \\
&+& 2\epsilon Tr \left[ \gamma_\mu \gamma\cdot p_1 \gamma\cdot \ell
\gamma_\nu \gamma\cdot p_2 \gamma\cdot \ell\right]
{1 \over {2p_1\cdot \ell}}\, {1 \over {2p_2\cdot \ell}};\nonumber \\
H^{(d)}_{\mu\nu} &=& -2 Tr \left[ \gamma_\mu \gamma\cdot (p_1 + \ell)
\gamma\cdot (p_2 + \ell) \gamma _\nu \gamma\cdot p_1 \gamma\cdot p_2\right]
{1 \over {2p_1 \cdot \ell}}\, {1 \over {2p_2\cdot \ell}}\nonumber \\
&+&
2\epsilon Tr \left[ \gamma_\mu \gamma\cdot \ell \gamma\cdot p_1 \gamma_\nu
\gamma\cdot \ell \gamma\cdot p_2\right]
{1 \over {2p_1\cdot \ell}}\, {1 \over {2 p_2 \cdot \ell}}.
\label{pp}
\end{eqnarray}
To obviate multiple repetition of a common factor, we temporarily omit the
overall coupling factor $e_q^2 (e\mu^\epsilon)^4$ that appears in
$H_{\mu\nu}$.  Function $H_1 = -g_{\mu\nu}\ H^{\mu\nu} = -g_{\mu\nu}
\sum\limits^{d}_{i=a} H^{(i)\mu\nu}$.  We obtain
\begin{equation}
H_1 = 8(1-\epsilon) \left\{ (1-\epsilon)
\left[ {{y_{1\ell}} \over {y_{2\ell}}} + {{y_{2\ell}}
       \over {y_{1\ell}}}\right]
+ {{2y_{12}} \over {y_{1\ell}\, y_{2\ell}} } -2\epsilon \right\}.
\label{qq}
\end{equation}
The dimensionless quantities $y_{1\ell}, y_{2\ell}$, and $y_{12}$
are defined by
\begin{eqnarray}
y_{i\ell}
&=& {{2p_i\cdot \ell} \over {q^2}}\ (i=1, 2); \nonumber \\
y_{12}
&=& {{2p_1\cdot p_2} \over {q^2}}.
\label{rr2}
\end{eqnarray}
We remark that $y_{12}+y_{1\ell}+y_{2\ell}=1$.
In evaluating $H_2 = -\left( k_\mu k_\nu /q^2\right) H^{\mu\nu}$, we
also make use of dimensionless quantities
$y_{1k}, y_{2k}$, and $y_{k\ell}$:
\begin{eqnarray}
y_{ik} &=& {{2p_i \cdot k} \over {q^2}}\ (i=1, 2); \nonumber \\
y_{k\ell} &=& {{2k\cdot \ell} \over {q^2}}.
\label{ss}
\end{eqnarray}
Because $k\cdot q = 0$,
\begin{equation}
y_{1k} + y_{2k} + y_{k\ell} = 0.
\label{tt}
\end{equation}
After some algebra we find
\begin{eqnarray}
H_2 &=&- 4 \left\{ (1-\epsilon)
\left[{{y_{1\ell}} \over {y_{2\ell}}} + {{y_{2\ell}} \over
      {y_{1\ell}}}\right]
+ {{2y_{12}} \over {y_{1\ell}\, y_{2\ell}} } -2\epsilon \right\}
\nonumber \\
&&+ {{4} \over {y_{1\ell}y_{2\ell}}}
   \left\{ y_{1k}^2 + y_{2k}^2 \right\}
 - {{4\epsilon} \over {y_{1\ell}y_{2\ell}}}
   \left\{ y_{k\ell}^2 \right\}.
\label{h2g}
\end{eqnarray}

Our next task is to integrate $H_1$ and $H_2$ over three-body phase
space in $n = 4-2\epsilon$ dimensions.
Since the momentum of the photon ($\ell$) is an observable,
and the momentum of either the quark ($p_1$) or antiquark ($p_2$)
can be fixed by the overall momentum conservation $\delta$-function in
the three-body phase space, we need to integrate over only $p_1$
or $p_2$.  In the following discussion, we let $p_2$ be fixed by the
$\delta$-function, and we integrate over $p_1$.
In the overall center of mass frame, as sketched in Fig.~\ref{fig8},
we take angle $\theta_\gamma$ to be the
polar angle of the $\gamma$
with respect to the $e^+e^-$ collision axis and
angle $\theta_{1\gamma}$ to be the angle
between the $\gamma$'s momentum $\ell$ and the quark momentum $p_1$.
The angle $\theta_x$ is the $n$-dimensional generalization of the
three-dimensional azimuthal angle $\phi$, defined through $p_1$ as
\begin{equation}
d\Omega_{n-2} (p_1) \equiv d\theta_{1\gamma} \sin^{n-3} \theta_{1\gamma}
d\theta_x \sin^{n-4} \theta_x d\Omega_{n-4} (p_1).
\label{vv}
\end{equation}

Having chosen the frame, we may reexpress the $y$ variables
in terms of observables and integration angles as follows:
\begin{eqnarray}
y_{k\ell}& =& - x_\gamma \cos \theta_\gamma\ , \nonumber \\
y_{1k}&=& -\left[\frac{y_{2\ell}y_{12}-y_{1\ell}}{x_\gamma}\right]
         \cos \theta_\gamma
         -\left[\frac{2\sqrt{y_{12}y_{1\ell}y_{2\ell}}}{x_\gamma}\right]
         \sin \theta_\gamma \cos \theta_x\ ; \nonumber \\
y_{2k}&=& -\left[\frac{y_{1\ell}y_{12}-y_{2\ell}}{x_\gamma}\right]
         \cos \theta_\gamma
         +\left[\frac{2\sqrt{y_{12}y_{1\ell}y_{2\ell}}}{x_\gamma}\right]
         \sin \theta_\gamma \cos \theta_x\ ;
\label{zz}
\end{eqnarray}
where $x_\gamma=2E_\gamma/\sqrt{s} (=y_{1\ell}+y_{2\ell})$.
In deriving $y_{1k}$ and $y_{2k}$,
we use the following identities
\begin{eqnarray}
\cos\theta_{1\gamma} &=&
\frac{y_{2\ell}y_{12}-y_{1\ell}}{x_1 x_\gamma}\ ;\nonumber \\
\sin\theta_{1\gamma} &=&
\frac{2\sqrt{y_{12}y_{1\ell}y_{2\ell}}}{x_1 x_\gamma}\ ,
\label{yy}
\end{eqnarray}
where $x_1=2E_1/\sqrt{s}$.

In the integration of $H_2$ over phase space, the integral over
$d\cos \theta_x$ is done from $\cos \theta_x = -1\ {\rm to}\ +1$.  The
expression for the three-body phase space, Eq.~(\ref{A.22}), is an even
function of $\cos \theta_x$.  Correspondingly, terms in $H_2$ that are odd
functions of $\cos \theta_x$  do not survive.
Because $H_2$ depends only on the square of the $y_{1k}$ and $y_{2k}$,
after eliminating all terms linear in $\cos\theta_x$, we find that
the only $\theta_x$
dependence in $H_2$ is $\cos^2\theta_x$.  We can integrate
over $\theta_x$ independent of other variables, or we can effectively
replace the $\cos^2\theta_x$ terms in $H_2$ by the average of
$\cos^2\theta_x$ in $n$-dimensions and eliminate the $\theta_x$
dependence in $H_2$ completely.

Given the average of $\cos^2\theta_x$ in $n$-dimensions,
Eq.~(\ref{A.25}), we obtain, effectively,
\begin{eqnarray}
y_{k\ell}^2 &=& x_\gamma^2 \cos^2\theta_\gamma\ ;\nonumber \\
y_{1k}^2&=& \left[\frac{y_{2\ell}y_{12}-y_{1\ell}}{x_\gamma}\right]^2
           \cos^2\theta_\gamma
          +\left(\frac{1}{1-\epsilon}\right)
           \left[\frac{2(y_{12}y_{1\ell}y_{2\ell})}{x_\gamma^2}\right]
           \sin^2\theta_\gamma\ ;\nonumber \\
y_{2k}^2&=& \left[\frac{y_{1\ell}y_{12}-y_{2\ell}}{x_\gamma}\right]^2
           \cos^2\theta_\gamma
          +\left(\frac{1}{1-\epsilon}\right)
           \left[\frac{2(y_{12}y_{1\ell}y_{2\ell})}{x_\gamma^2}\right]
           \sin^2\theta_\gamma \ ;
\label{xx}
\end{eqnarray}
where the factor $1/(1-\epsilon)$ is from the average of
$\cos^2\theta_x$.  Substituting the
above expressions into Eq.~(\ref{h2g}),
and combining with $H_1$ in Eq.~(\ref{qq}), we obtain,
\begin{eqnarray}
{{1} \over {4}} \left( H_1+H_2^{eff} \right)
&=&\left( 1 + \cos^2 \theta_\gamma - 2\epsilon\right)
   \left[ (1-\epsilon)
        \left(\frac{y_{1\ell}}{y_{2\ell}}
             +\frac{y_{2\ell}}{y_{1\ell}}\right)
       +2\left(\frac{y_{12}}{y_{1\ell}y_{2\ell}}-\epsilon\right) \right]
\nonumber \\
&+& \left( 1-3\cos^2\theta_\gamma\right)
   \left[\frac{4\, y_{12}}{x_\gamma^2}\right]\ \nonumber \\
&+& \left(\frac{\epsilon}{1-\epsilon}\right)
   \left(1-\cos^2\theta_\gamma\right)
   \left[\frac{4\, y_{12}}{x_\gamma^2} \right]\ ,
\label{h1h2}
\end{eqnarray}
where the superscript ``{\scriptsize\it eff}''
indicates that we have replaced
$\cos^2\theta_x$ by its average in $n$-dimensions.
The two $\delta$-functions in the three-particle phase space,
$dPS^{(3)}$, provide the following identities
\begin{eqnarray}
y_{12} &=& 1-x_\gamma\ ; \nonumber \\
y_{2\ell} &=& x_\gamma - y_{1\ell}\ .
\label{yid}
\end{eqnarray}
Introducing $\hat{y}_{1\ell} = y_{1\ell}/x_\gamma$, and
substituting these identities into Eq.~(\ref{h1h2}), we derive
\begin{eqnarray}
{{1} \over {4}} \left( H_1+H_2^{eff} \right)
&=& \left( 1 + \cos^2 \theta_\gamma - 2\epsilon\right)
   \left[ {{1+(1-x_\gamma)^2} \over {x^2_\gamma}}\right]
   \left( {{1} \over {\hat{y}_{1\ell}}}
  +{{1} \over {1-\hat{y}_{1\ell}}}\right)\nonumber \\
&+& \left( 1 + \cos^2\theta_\gamma - 2\epsilon\right)
   \left[ -2-\epsilon\left( {{1}\over{\hat{y}_{1\ell}}}
  +{{1}\over {1-\hat{y}_{1\ell}}}\right)\right]\nonumber \\
&+& \left( 1-3\cos^2\theta_\gamma\right)
   \left[{{4(1-x_\gamma)} \over {x^2_\gamma}}\right] \nonumber \\
&+& \left(\frac{\epsilon}{1-\epsilon}\right)
   \left(1-\cos^2\theta_\gamma\right)
   \left[{{4(1-x_\gamma)}\over{x^2_\gamma}} \right]\ .
\label{ggg}
\end{eqnarray}
The last term vanishes as $\epsilon\rightarrow 0$.

Combining Eqs.~(\ref{oo}) and (\ref{ggg}), and integrating over
$d\hat{y}_{1\ell}$, we can derive the partonic cross section
$d\sigma^{(1)}_{e^+e^-\rightarrow\gamma X}$.  The limits of the
$d\hat{y}_{1\ell}$ integration are from 0 to 1.
The integrals over $d\hat{y}_{1\ell}$ for $(H_1+H_2^{eff})/4$ may be
expressed in terms of
\begin{equation}
I_{n,m} \equiv \int^1_0\ d\hat{y}_{1\ell}\ \hat{y}^{n-\epsilon}_{1\ell}
\left( 1- \hat{y}_{1\ell}\right)^{m-\epsilon}.
\label{hhh}
\end{equation}
Examining Eq.~(\ref{ggg}), we need only $I_{0,0}$ and $I_{-1,0}
\left( = I_{0, -1}\right)$:
\begin{equation}
I_{0,0} = B(1-\epsilon, 1-\epsilon) =
{\left({\Gamma(1-\epsilon)}\right)^2 \over {\Gamma(2-2\epsilon)}};
\label{iii}
\end{equation}
\begin{equation}
I_{-1,0} = B(-\epsilon, 1-\epsilon) = \left( {{1} \over {-\epsilon}}\right)
{{\left( \Gamma(1-\epsilon)\right)^2} \over {\Gamma(1-2\epsilon)}}.
\label{jjj}
\end{equation}
For small $\epsilon,\ I_{0,0} = 1 + O(\epsilon)$, and
$I_{-1,0} = - {{1}\over{\epsilon}} + O(\epsilon)$.

After performing the integration over $d\hat{y}_{1\ell}$,
we expand the right-hand side of Eq.~(\ref{oo}) in a power series
in $\epsilon$, keeping only the singular term
proportional to $(1/\epsilon)$ and the terms independent of $\epsilon$.
(Terms of $O(\epsilon^m),\ m \geq 1$, vanish in the physical limit of four
dimensions $(n = 4-2\epsilon)$). We obtain
\newpage
\begin{eqnarray}
E_\gamma {{d\sigma^{(1)}_{e^+e^- \rightarrow \gamma X}} \over {d^3\ell}}
&=&\ 2 \sum_{q} \left[{{2}\over{s}} F^{PC}_{q} (s)\right]
     \left[\alpha^2_{em}N_c
           \left({{4\pi\mu^2} \over
                 {(s/4)\sin^2\theta_\gamma}}\right)^\epsilon
           {{1} \over{\Gamma(1-\epsilon)}} \right]
      \left( 1 + \cos^2 \theta_\gamma -2\epsilon\right) \nonumber \\
&& \times {{1}\over {x_\gamma}}
      \Bigg\{ e^2_q\ {{\alpha_{em}}\over{2\pi}}
      \left[ {{1+(1-x_\gamma)^2}\over{x_\gamma}}\right]\Bigg\}
      \left(-{{1}\over{\epsilon}}\right)\nonumber \\
&+&\ 2\sum_{q}\left[{{2}\over {s}}\ F^{PC}_{q} (s)\right]
    \left[ \alpha^2_{em} N_c {{1}\over {x_\gamma}} \right]
    e^2_q \left( {{\alpha_{em}}\over {2\pi}}\right)\nonumber \\
&& \times \Bigg\{ (1+\cos^2\theta_\gamma)
    \left[ {{1+(1-x_\gamma)^2}\over{x_\gamma}}\right]
    \left[ \ell n \left( s/\mu^2_{\overline{\rm MS}}\right)
          +\ell n\left(x^2_\gamma \left(1-x_\gamma\right)\right) \right]
\nonumber \\
&& +\left( 1-3\cos^2\theta_\gamma\right)
   \left[{{2(1-x_\gamma)}\over{x_\gamma}}\right] \Bigg\}\ .
\label{kkk}
\end{eqnarray}
In deriving Eq.~(\ref{kkk}), we included the overall factor for
coupling constants, $e_q^2\ (e\mu^\epsilon)^4$; and used the expansion
$\Gamma(1-\epsilon) \simeq 1 + \epsilon \gamma_E$, where $\gamma_E$ is
Euler's constant, and the usual modified minimal subtraction scale
\begin{equation}
\mu^2_{\overline{\rm MS}} \equiv \mu^2 4\pi e^{-\gamma_E}.
\label{lll}
\end{equation}

The $(1/\epsilon)$ singularity in Eq.~(\ref{kkk}) represents the
quark-photon
collinear singularity.  This singular term is expected to be cancelled
by subtraction terms defined in Eq.~(\ref{kk}).  By evaluating the
diagram sketched in Fig.~\ref{fig9}, we obtain the one-loop
quark-to-photon fragmentation function
\begin{equation}
D^{(1)}_{q\rightarrow \gamma} (z) = D^{(1)}_{\bar{q}\rightarrow\gamma}(z)
= e^2_q\ {{\alpha_{em}} \over {2\pi}}\
\left[ {{1 + (1-z)^2} \over {z}}\right]
\left( {{1} \over {-\epsilon}}\right)\ ,
\label{mmm}
\end{equation}
where we keep only the $1/\epsilon$ pole term because we work in the
$\overline{\rm MS}$ factorization scheme.  Using the fact that
$D^{(1)}_{\bar{q}\rightarrow\gamma}(z)=
 D^{(1)}_{q\rightarrow\gamma}(z)$, and
comparing Eq.~(\ref{kkk}) with
Eqs.~(\ref{dd}) and (\ref{mmm}), we observe that
the divergent first term in Eq.~(\ref{kkk}) is cancelled exactly by
the subtraction terms defined in Eq.~(\ref{kk}), in accord with the
pQCD factorization theorem.
Using Eq.~(\ref{kk}), we obtain the finite
$O(\alpha_{em})$ hard-scattering cross section
\begin{eqnarray}
E_\gamma
{{d\hat{\sigma}^{(1)}_{e^+e^-\rightarrow\gamma X}} \over {d^3\ell}}
&=&\ 2 \sum_q \left[ {{2} \over {s}} F^{PC}_q\, (s)\right]
      \left[ \alpha^2_{em} N_c {{1} \over {x_\gamma}} \right]
      e_q^2 \left( {{\alpha_{em}} \over {2\pi}}\right) \nonumber \\
&&\times \Bigg\{ \left(1+\cos^2 \theta_\gamma\right)
\left[ {{1+(1-x_\gamma)^2} \over {x_\gamma}}\right]
\left[ \ell n \left(s/\mu^2_{\overline{\rm MS}}\right)
     + \ell n \left( x^2_\gamma \left(1-x_\gamma\right) \right) \right]
\nonumber \\
&&+ (1-3 \cos^2 \theta_\gamma )
   \left[{{2(1-x_\gamma)} \over {x_\gamma}} \right]\Bigg\}.
\label{nnn}
\end{eqnarray}

We remark that the angular dependence of the $O(\alpha_{em})$
hard-scattering cross section has two components, one proportional
to $(1 + \cos^2\theta)$, familiar from the lowest order expression,
and a second piece proportional to $(1-3  \cos^2 \theta_\gamma)$.
If one integrates over $\cos\theta_\gamma$, the second piece vanishes.
We note, however, that the piece proportional to
$(1-3 \cos^2 \theta_\gamma)$ changes the predicted angular
dependence from the often assumed form $(1 + \cos^2\theta)$.
The difference means that it would not be correct to assume a
$(1 + \cos^2\theta)$ dependence when attempting to
correct an integrated cross section for unobserved events near,
for example, the incident $e^+e^-$ beam direction,
$\cos\theta_\gamma = \pm 1$.

\subsection{Derivation of $\hat{\sigma}^{(1)}_{e^+e^- \rightarrow gX}$}
\label{subsec:3b}

The finite hard-scattering cross section
$\hat{\sigma}^{(1)}_{e^+e^-\rightarrow gX}$ to first order in $\alpha_s$
may be obtained directly from Eq.~(\ref{nnn}) after three replacements:
$x_\gamma \rightarrow x_g;\ N_c \rightarrow N_c C_F$; and $e^2e^2_q$
of the final photon emission vertex by $g^2 = 4 \pi \alpha_s$.
\begin{eqnarray}
E_g {{d\hat{\sigma}^{(1)}_{e^+e^-\rightarrow gX}} \over {d^3p_g}}
&=&\ 2 \sum_q \left[ {{2}\over {s}}\, F^{PC}_q\, (s)\right]
  \left[\alpha^2_{em}\, N_c\, {{1}\over {x_g}}\right]
  C_F\left( {{\alpha_s}\over {2\pi}}\right) \nonumber \\
&&\times \Bigg\{ \left( 1 + \cos^2\theta_g \right)
  \left[ {{1+(1-x_g)^2} \over {x_g}}\right]
  \left[ \ell n \left( s/\mu^2_{\overline{\rm MS}}\right)
 +\ell n \left(x^2_g\left(1-x_g\right)\right) \right] \nonumber \\
&&+ \left( 1-3\cos^2 \theta_g \right)
 \left[{{2(1-x_g)} \over {x_g}}\right]\Bigg\}.
\label{ooo}
\end{eqnarray}
In Eq.~(\ref{ooo}), $x_g = 2E_g/\sqrt{s}$; $C_F = {4\over 3}$, and
$N_c=3$.

The contribution $O(\alpha_s)$ to the inclusive yield
$e^+e^-\rightarrow\gamma X$ via gluon fragmentation is therefore
\begin{equation}
E_\gamma {{d\sigma^{(1)}_{e^+e^-\rightarrow gX\rightarrow \gamma X}}
\over {d^3\ell}} = \int^1_{x_\gamma}\, {{dz} \over {z}}
\left[ E_g {{d\hat{\sigma}_{e^+e^-\rightarrow gX}^{(1)}} \over {d^3p_g}}\,
\left( x_g = {{x_\gamma} \over {z}}\right)\right]\,
{{D_{g\rightarrow\gamma}(z, \mu^2_{\overline{\rm MS}})} \over {z}}
\label{ppp}
\end{equation}
with $x_\gamma = 2E_\gamma/\sqrt{s}$.  Because the $g\rightarrow \gamma$
fragmentation process is collinear, $\theta_g = \theta_\gamma$.

\subsection{Derivation of $\hat{\sigma}^{(1)}_{e^+e^- \rightarrow qX}$}
\label{subsec:3c}

In this section we present our explicit derivation of the finite
hard-scattering cross section $E_q$
$\hat{\sigma}^{(1)}_{e^+e^- \rightarrow qX}/d^3p_q$
to first order in $\alpha_s$.  As sketched in Fig.~\ref{fig6},
both real gluon
emission and virtual gluon exchange graphs contribute.
The real emission diagrams have both infrared and
collinear divergences.  The infrared divergence is cancelled by
contributions from the virtual diagrams, while the collinear divergence
is cancelled by the subtraction term defined in Eq.~(\ref{nn}).

The real emission diagrams can be treated easily in the same way as
$d\sigma^{(1)}_{e^+e^- \rightarrow \gamma X}/d^3\ell$ in
Section~\ref{subsec:3a}.
Except for the replacement of a photon by a gluon, the hadronic tensor
$H_{\mu\nu}$ obtained from the gluon emission diagrams in
Fig.~\ref{fig6}a is
identical to that computed in Section~\ref{subsec:3a}  for
$e^+e^- \rightarrow q\bar{q} \gamma$.  Thus, we may employ
our previous expressions for $H_1$ and $H_2$ again
but with the replacement of subscript ``$\ell$'' in
Eqs. (\ref{qq}) and (\ref{h2g}) by subscript ``3'', since $p_3$
is our momentum label for the gluon.  Because the quark is now the
fragmenting particle (i.e., effectively the ``observed'' particle),
the $y_{ik}^2$ variables with $i=1,2,3$ in $H_2$ are no longer
those in Eq.~(\ref{xx}).  Instead, we now have
\begin{eqnarray}
y_{1k}^2 &=& x_1^2 \cos^2\theta_1\ ;\nonumber \\
y_{2k}^2&=& \left[\frac{y_{13}y_{23}-y_{12}}{x_1}\right]^2
           \cos^2\theta_1
          +\left(\frac{1}{1-\epsilon}\right)
           \left[\frac{2(y_{12}y_{13}y_{23})}{x_1^2}\right]
           \sin^2\theta_1\ ;\nonumber \\
&& \mbox{and} \nonumber \\
y_{3k}^2&=& \left[\frac{y_{12}y_{23}-y_{13}}{x_1}\right]^2
           \cos^2\theta_1
          +\left(\frac{1}{1-\epsilon}\right)
           \left[\frac{2(y_{12}y_{13}y_{23})}{x_1^2}\right]
           \sin^2\theta_1\ .
\label{xxq}
\end{eqnarray}
In Eq.~(\ref{xxq}), $\theta_1$ is the scattering angle of the quark, and
subscript ``3'' indicates the gluon of momentum $p_3$.  We
dropped all terms linear in $\cos\theta_x$, and replaced
$\cos^2\theta_x$ by its average value in $n$-dimensions.  Substituting
these $y_{ik}^2$ with $i=1,2,3$ into Eq.~(\ref{h2g}), we derive
\begin{eqnarray}
{{1} \over {4}} \left( H_1+H_2^{eff} \right)
&= &\left( 1 + \cos^2 \theta_1 - 2\epsilon\right)
   \left[ (1-\epsilon)
        \left(\frac{y_{13}}{y_{23}}
             +\frac{y_{23}}{y_{13}}\right)
       +2\left(\frac{y_{12}}{y_{13}y_{23}}-\epsilon\right) \right]
\nonumber \\
&+& \left( 1-3\cos^2\theta_1\right)
   \left[\frac{2\, y_{12}}{x_1^2}\right]\ \nonumber \\
&&+\ \epsilon\ \cos^2\theta_1
   \left[\frac{4\, y_{12}}{x^2_1}\right]\ ,
\label{h12q}
\end{eqnarray}
where the last term again vanishes as $\epsilon\rightarrow 0$.
In analogy to Eq.~(\ref{yid}), the useful identities here are
\newpage
\begin{eqnarray}
y_{23} &=& 1-x_1\ ; \nonumber \\
y_{12} &=& x_1 - y_{13}\ .
\label{yidq}
\end{eqnarray}
Using these identities, we reexpress Eq.~(\ref{h12q}) in terms of
$x_1$ and $y_{13}$
\begin{eqnarray}
{{1} \over {4}} \left( H_1+H_2^{eff} \right)
&=& \left( 1 + \cos^2 \theta_1 - 2\epsilon\right)
   \Bigg\{ \left[{{1+x_1^2} \over {1-x_1}}\right]
           {{1} \over {y_{13}}}
          +{{y_{13}} \over {1-x_1}}
          -{{2} \over {1-x_1}} \nonumber \\
&&\mbox{\hskip 1.6in} -\epsilon\left[
   {{1-x_1}\over{y_{13}}}
  +{{y_{13}}\over {1-x_1}} + 2 \right] \Bigg\} \nonumber \\
&+& \left( 1-3\cos^2\theta_1\right)
   \left[ \frac{2}{x_1}\left(1-\frac{y_{13}}{x_1}\right) \right]\ ,
\label{h12q2}
\end{eqnarray}
where we dropped the last term in Eq.~(\ref{h12q}).
Introducing the overall coupling factor $(e\mu^\epsilon)^2
(g\mu^\epsilon)^2$ and color factor $N_c C_F$,
and combining with the three particle final state
phase space $dPS^{(3)}$, Eq.~(\ref{A.27}),
we express the contribution of real gluon
emission as
\begin{eqnarray}
E_1 {{d\sigma^{(R)}_{e^+e^-\rightarrow qX}} \over {d^3p_1}}
&=& \left[ {{2}\over {s}}\, F^{PC} (s)\right]
   \left[ \alpha^2_{em}\, N_c
   \left( {{4\pi \mu^2} \over {(s/4) \sin^2\theta_1}}\right)^\epsilon
          {{1} \over {\Gamma(1-\epsilon)}} \right]\nonumber \\
&\times & C_F \left( {{\alpha_s} \over {2\pi}}\right)
\left[ \left({{4\pi \mu^2} \over {s}}\right)^\epsilon
             {{1} \over {\Gamma(1-\epsilon)}} \right]
             {{\delta\left( x_1-(1-y_{23})\right)} \over {x_1}}\nonumber \\
&\times &{{1}\over {4}} \left[ H_1 + H_2^{eff} \right]
        {{dy_{12}}\over{y^\epsilon_{12}}}\,
        {{dy_{13}}\over{y^\epsilon_{13}}}\,
        {{dy_{23}}\over{y^\epsilon_{23}}}\,
        \delta\left( 1-y_{12}-y_{13}-y_{23}\right)\ ,
\label{qqq}
\end{eqnarray}
where superscript $(R)$ stands for the real emission.
Using the two $\delta$-functions to fix $y_{23}$ and $y_{12}$,
inserting $H_1+H_2^{eff}$ from Eq.~(\ref{h12q2}),
and integrating $dy_{13}$ from 0 to $x_1$, we derive
\begin{eqnarray}
E_1 {{d\sigma^{(R)}_{e^+e^-\rightarrow qX}} \over {d^3p_1}}
&=& \left[ {{2}\over {s}}\, F^{PC} (s)\right]
   \left[ \alpha^2_{em}\, N_c
   \left( {{4\pi \mu^2} \over {(s/4) \sin^2\theta_1}}\right)^\epsilon
          {{1} \over {\Gamma(1-\epsilon)}} \right]\nonumber \\
&\times &C_F \left( {{\alpha_s} \over {2\pi}}\right)
\left[ \left({{4\pi \mu^2} \over {s}}\right)^\epsilon
             {{1} \over {\Gamma(1-\epsilon)}} \right]
             \left({{1} \over {x_1}}\right)
             \frac{\Gamma(1-\epsilon)^2}{\Gamma(1-2\epsilon)} \nonumber \\
&\times &\Bigg\{
 \left( 1 + \cos^2 \theta_1 - 2\epsilon\right)\Bigg[
      \left(\frac{1+x_1^2}{(1-x_1)_+}+\frac{3}{2}\delta(1-x_1)\right)
      \left(\frac{1}{-\epsilon}\right) \nonumber \\
&&\quad\quad\quad
       +\left(\frac{1+x_1^2}{1-x_1}\right)\ell n\left(x_1^2\right)
       +\left(1+x_1^2\right)\left(\frac{\ell n(1-x_1)}{1-x_1}\right)_+
       -\frac{3}{2}\left(\frac{1}{1-x_1}\right)_+ \nonumber \\
&&\quad\quad\quad
       +\delta(1-x_1)\left(
       \frac{2}{\epsilon^2}+\frac{3}{\epsilon}+\frac{7}{2} \right)
      -\frac{1}{2}\left(3x_1-5\right) \Bigg] \nonumber \\
&+& \left(1-3\cos^2\theta_1 \right) \Bigg\} \ .
\label{realg}
\end{eqnarray}
\newpage
\noindent The ``+'' prescription is defined as usual
\begin{equation}
\left(\frac{1}{1-x_1}\right)_+ \equiv
\left(\frac{1}{1-x_1}\right)
-\delta(1-x_1)\int_0^1 dz \left(\frac{1}{1-z}\right) \ .
\label{plus}
\end{equation}
The right-hand-side of Eq.~(\ref{realg}) is formally divergent as
$\epsilon\rightarrow 0$.  The $1/\epsilon$ poles in $n$-dimensions
represent the infrared divergence, when the gluon momentum goes to zero,
and/or a collinear divergence, when the gluon momentum is parallel to
that of the
fragmenting quark.  As we show below, the infrared divergence is
cancelled by the infrared divergence of the virtual diagrams, sketched
in Fig.~\ref{fig6}b.

The contribution of the virtual diagrams
results from the interference of the one-loop vertex and self-energy
diagrams with the leading order tree diagram.  As for the leading
order contribution, the virtual diagrams, sketched in
Fig.~\ref{fig6}b, have a
two-particle final state phase space.  Therefore, the contribution from
the virtual diagrams has the same kinematical structure
and angular dependence
as the leading order contribution, discussed in Section~\ref{subsec:2c}.
It is proportional to $\delta(1-x_1)$, and, consequently, the virtual
contribution cancels only the $1/\epsilon$ poles associated with the
$\delta(1-x_1)$ terms in Eq.~(\ref{realg}).
The subtraction terms in Eq.~(\ref{nn}) cancel the final state
collinear poles that appear in the contribution of real gluon emission.

Beginning with the virtual exchange diagrams in Fig.~\ref{fig6}b,
we evaluate the
one-loop vertex correction in $n$-dimension, and combine it with the
lowest order tree diagram to form the first order virtual contribution.
We derive
\begin{eqnarray}
E_1 {{d\sigma^{(V)}_{e^+e^-\rightarrow qX}} \over {d^3p_1}}
&=& \left[ {{2}\over {s}}\, F^{PC} (s)\right]
   \left[ \alpha^2_{em}\, N_c
   \left( {{4\pi \mu^2} \over {(s/4) \sin^2\theta_1}}\right)^\epsilon
          {{1} \over {\Gamma(1-\epsilon)}} \right]\nonumber \\
&\times &C_F \left( {{\alpha_s} \over {2\pi}}\right)
\left[ \left({{4\pi \mu^2} \over {s}}\right)^\epsilon
             {{1} \over {\Gamma(1-\epsilon)}} \right]
             \left({{1} \over {x_1}}\right)
             \frac{\Gamma(1-\epsilon)^3\Gamma(1+\epsilon)}
                  {\Gamma(1-2\epsilon)} \nonumber \\
&\times &\Bigg\{
 \left( 1 + \cos^2 \theta_1 - 2\epsilon\right)
       \delta(1-x_1)
       \left[-\frac{2}{\epsilon^2}-\frac{3}{\epsilon}
       +\left(\pi^2-8\right)\right] \Bigg\} \ ,
\label{virtual}
\end{eqnarray}
where the superscript (V) stands for the virtual contribution.  After
adding the real and virtual contributions, Eqs.~(\ref{realg})
and (\ref{virtual}), we obtain the cross
section for $e^+e^-\rightarrow qX$ at order $O(\alpha_s)$,
\newpage
\begin{eqnarray}
E_1 {{d\sigma^{(1)}_{e^+e^-\rightarrow qX}} \over {d^3p_1}}
&=&\ \left[ {{2}\over {s}}\, F^{PC} (s)\right]
    \left[ \alpha^2_{em}\, N_c
    \left( {{4\pi \mu^2} \over {(s/4) \sin^2\theta_1}}\right)^\epsilon
          {{1} \over {\Gamma(1-\epsilon)}} \right]\nonumber \\
&&\times C_F \left( {{\alpha_s} \over {2\pi}}\right)
      {{1} \over {x_1}} \Bigg\{
      \left( 1 + \cos^2 \theta_1 - 2\epsilon\right)
      \left[\frac{1+x_1^2}{(1-x_1)_+}+\frac{3}{2}\delta(1-x_1)\right]
      \left(\frac{1}{-\epsilon}\right) \Bigg\} \nonumber \\
&+&\ \left[ {{2}\over {s}}\, F^{PC} (s)\right]
   \left[ \alpha^2_{em}\, N_c\, {{1} \over {x_1}} \right]
          C_F \left( {{\alpha_s} \over {2\pi}}\right) \nonumber \\
&&\times\Bigg\{ (1+\cos^2\theta_1)
   \Bigg[\left(\frac{1+x_1^2}{(1-x_1)_+}+\frac{3}{2}\delta(1-x_1)\right)
               \ell n\left(\frac{s}{\mu^2_{\overline{\rm MS}}}\right)
\nonumber \\
&&\quad\quad + \left(\frac{1+x_1^2}{1-x_1}\right)\ell n\left(x_1^2\right)
       + \left(1+x_1^2\right)\left(\frac{\ell n(1-x_1)}{1-x_1}\right)_+
       - \frac{3}{2}\left(\frac{1}{1-x_1}\right)_+ \nonumber \\
&&\quad\quad + \delta(1-x_1)\left(\frac{2\pi^2}{3}-\frac{9}{2}\right)
       - \frac{1}{2}\left(3x_1-5\right) \Bigg] \nonumber \\
&& + \left(1-3\cos^2\theta_1\right) \Bigg\} \ .
\label{rv}
\end{eqnarray}
As is evident from the $1/\epsilon$ terms,
this cross section is divergent as $\epsilon\rightarrow 0$,
a reflection of the fact that a cross section for producing a
massless quark is an infrared sensitive quantity, not
perturbatively calculable.

According to the pQCD factorization theorem, the short-distance
hard-scattering cross sections, defined in Eq.~(\ref{a}), are infrared
safe quantities.  Beyond the Born level, the short-distance parts,
$\hat{\sigma}_{e^+e^-\rightarrow cX}$, are not the same as the partonic
cross sections $\sigma_{e^+e^-\rightarrow cX}$ for fragmenting parton
$c$. Following Eq.~(\ref{nn}), in order to derive the short-distance
hard-scattering cross section $\hat{\sigma}^{(1)}_{e^+e^-\rightarrow qX}$,
we must first calculate the one-loop perturbative
fragmentation function $D^{(1)}_{q\rightarrow q}$.
Feynman diagrams for $D^{(1)}_{q\rightarrow q}$ are sketched in
Fig.~\ref{fig10}.
These diagrams are evaluated in the same way as one evaluates
parton-level parton distributions \cite{CQ}, and we obtain
\begin{equation}
D^{(1)}_{q\rightarrow q} (x_1) =
C_F \left({{\alpha_{s}} \over {2\pi}}\right)
\left[ \frac{1 + x_1^2}{(1-x_1)_+}
+\frac{3}{2}\delta(1-x_1)\right]
\left( {{1} \over {-\epsilon}}\right)\ ,
\label{dzqq}
\end{equation}
where the ``+'' prescription is defined in Eq.~(\ref{plus}).

Using Eq.~(\ref{nn}), the lowest order cross section for
$e^+e^-\rightarrow qX$, Eq.~(\ref{dd}),
and the one-loop quark fragmentation function
$D^{(1)}_{q\rightarrow q}$, Eq.~(\ref{dzqq}),
we derive the short-distance hard-scattering cross section
\begin{eqnarray}
E_1 \frac{d\hat{\sigma}^{(1)}_{e^+e^-\rightarrow qX}}{d^3p_1}
&= & \, \left[\frac{2}{s}F^{PC}_{q}(s)\right]
               \left[\alpha_{em}^2N_c \frac{1}{x_1} \right]
           C_F \left(\frac{\alpha_{s}}{2\pi}\right) \nonumber \\
&\times & \Bigg\{(1+\cos^2\theta_\gamma)
   \Bigg[\left(\frac{1+x_1^2}{(1-x_1)_+}+\frac{3}{2}\delta(1-x_1)\right)
               \ell n\left(\frac{s}{\mu^2_{\overline{\rm MS}}}\right)
\nonumber \\
&&\quad\quad + \left(\frac{1+x_1^2}{1-x_1}\right)\ell n\left(x_1^2\right)
       + \left(1+x_1^2\right)\left(\frac{\ell n(1-x_1)}{1-x_1}\right)_+
       - \frac{3}{2}\left(\frac{1}{1-x_1}\right)_+ \nonumber \\
&&\quad\quad + \delta(1-x_1)\left(\frac{2\pi^2}{3}-\frac{9}{2}\right)
       - \frac{1}{2}\left(3x_1-5\right) \Bigg] \nonumber \\
&&+ \left(1-3\cos^2\theta_\gamma\right) \Bigg\} .
\label{hardq}
\end{eqnarray}
We set $\theta_\gamma=\theta_1$ based on the assumption of
collinear fragmentation from quark to photon.  As expected, the
hard-scattering cross section is infrared insensitive.  The
$O(\alpha_s)$ quark fragmentation contribution to $e^+e^-\rightarrow
\gamma X$ is
\begin{equation}
E_\gamma \frac{d\sigma^{(1)}_{e^+e^-\rightarrow qX\rightarrow \gamma X}}
              {d^3\ell}
= \sum_q \int^1_{x_\gamma}\, \frac{dz}{z} \left[
E_1 \frac{d\hat{\sigma}_{e^+e^-\rightarrow qX}^{(1)}}{d^3p_1}
\left( x_1 = \frac{x_\gamma}{z}\right)\right]\,
\frac{D_{q\rightarrow\gamma}(z, \mu^2_{\overline{\rm MS}})}{z} \ .
\label{fragq}
\end{equation}

Our derivation shows that the short-distance hard-scattering
cross section for antiquark fragmentation to a photon is
the same as that for quark fragmentation.  Consequently, the $O(\alpha_s)$
antiquark fragmentation contribution to $e^+e^-\rightarrow \gamma X$
is the same as that given in Eq.~(\ref{fragq}).

\section{Numerical Results and Discussion}
\label{sec:result}

In this section we present and discuss explicit
numerical evaluations of the
inclusive prompt photon cross sections derived in this paper.
We provide results at $e^+e^-$ center-of-mass energies
$\sqrt{s} = $ 10 GeV, 58 GeV, and 91 GeV appropriate for
experimental investigations underway at Cornell, KEK, SLAC, and CERN.
In our figures, we display the variation of the inclusive
yield with photon energy $E_\gamma$ and scattering angle $\theta _\gamma$,
where $\theta_\gamma$ is the angle of the photon with respect
to the $e^+e^-$ collision axis.  We also show the dependence of
cross sections on the choice of renormalization scale $\mu$.

The cross sections we evaluate are those derived in the text:
Eqs.~(\ref{ee}), (\ref{nnn}), (\ref{ppp}), and (\ref{fragq}).
They are assembled here for convenience of comparison.
The lowest order inclusive cross section is
\begin{equation}
E_\gamma {{d\sigma^{incl}_{e^+e^-\rightarrow \gamma X}} \over d^3 \ell}
= 2\sum_q \left[ {{2} \over {s}} F^{PC}_q (s)\right] \alpha^2_{em}(s)
N_c (1 + \cos^2\theta_\gamma) {{1} \over {x_\gamma}}
D_{q \rightarrow \gamma} (x_\gamma, \mu^2_F).
\label{aaaa}
\end{equation}
The finite $O(\alpha_{em})$ hard-scattering cross section is
\begin{eqnarray}
E_\gamma
{{d\hat{\sigma}^{(1)}_{e^+e^-\rightarrow\gamma X}} \over {d^3\ell}}
&=&\ 2 \sum_q \left[ {{2} \over {s}} F^{PC}_q\, (s)\right]
      \left[ \alpha^2_{em}(s) N_c {{1} \over {x_\gamma}} \right]
      e_q^2 \left( {{\alpha_{em}(\mu^2_F)} \over {2\pi}}\right)
\nonumber \\
&&\times \Bigg\{ \left(1+\cos^2 \theta_\gamma\right)
\left[ {{1+(1-x_\gamma)^2} \over {x_\gamma}}\right]
\left[ \ell n \left(s/\mu^2_F\right)
     + \ell n \left( x^2_\gamma \left(1-x_\gamma\right) \right) \right]
\nonumber \\
&&+ (1-3 \cos^2 \theta_\gamma )
   \left[{{2(1-x_\gamma)} \over {x_\gamma}} \right]\Bigg\}\ .
\label{bbbb}
\end{eqnarray}

\noindent The $O(\alpha_s)$ contribution to the inclusive yield
$e^+e^-\rightarrow \gamma X$ via gluon fragmentation is
\begin{equation}
E_\gamma {{d\sigma^{(1)}_{e^+e^-\rightarrow gX\rightarrow \gamma X}}
\over {d^3\ell}} = \int^1_{x_\gamma}\, {{dz} \over {z}}
\left[ E_g {{d\hat{\sigma}_{e^+e^-\rightarrow gX}^{(1)}} \over {d^3p_g}}\,
\left( x_g = {{x_\gamma} \over {z}}\right)\right]\,
{{D_{g\rightarrow\gamma}(z, \mu^2_F)} \over {z}}
\label{cccc}
\end{equation}
with
\begin{eqnarray}
E_g {{d\hat{\sigma}^{(1)}_{e^+e^-\rightarrow gX}} \over {d^3p_g}}
&=&\ 2 \sum_q \left[ {{2}\over {s}}\, F^{PC}_q\, (s)\right]
  \left[\alpha^2_{em}(s)\, N_c\, {{1}\over {x_g}}\right]
  C_F\left( {{\alpha_s(\mu^2_F)}\over {2\pi}}\right) \nonumber \\
&&\times \Bigg\{ \left( 1 + \cos^2\theta_\gamma \right)
  \left[ {{1+(1-x_g)^2} \over {x_g}}\right]
  \left[ \ell n \left( s/\mu^2_F\right)
 +\ell n \left(x^2_g\left(1-x_g\right)\right) \right]\nonumber \\
&&+ \left( 1-3\cos^2 \theta_\gamma \right)
 \left[{{2(1-x_g)} \over {x_g}}\right]\Bigg\}\ .
\label{dddd}
\end{eqnarray}
We choose the renormalization scale $\mu$ in
$\alpha_s(\mu^2)$ to be the same as the fragmentation scale
$\mu_F$ in $D_{g\rightarrow\gamma}(z,\mu^2_F)$.
The $O(\alpha_s)$ contribution to the inclusive yield
$e^+e^-\rightarrow \gamma X$ via quark fragmentation is
\begin{equation}
E_\gamma \frac{d\sigma^{(1)}_{e^+e^-\rightarrow qX\rightarrow \gamma X}}
              {d^3\ell}
= \sum_q \int^1_{x_\gamma}\, \frac{dz}{z} \left[
E_1 \frac{d\hat{\sigma}_{e^+e^-\rightarrow qX}^{(1)}}{d^3p_1}
\left( x_1 = \frac{x_\gamma}{z}\right)\right]\,
\frac{D_{q\rightarrow\gamma}(z,\mu^2_F)}{z} \ .
\label{eeee}
\end{equation}
\newpage
\noindent with
\begin{eqnarray}
E_1 \frac{d\hat{\sigma}^{(1)}_{e^+e^-\rightarrow qX}}{d^3p_1}
&= & \, \left[\frac{2}{s}F^{PC}_{q}(s)\right]
               \left[\alpha_{em}^2(s) N_c \frac{1}{x_1} \right]
           C_F \left(\frac{\alpha_{s}(\mu^2_F)}{2\pi}\right) \nonumber \\
&\times & \Bigg\{(1+\cos^2\theta_\gamma)
   \Bigg[\left(\frac{1+x_1^2}{(1-x_1)_+}+\frac{3}{2}\delta(1-x_1)\right)
               \ell n\left(\frac{s}{\mu^2_F}\right) \nonumber \\
&&\quad\quad + \left(\frac{1+x_1^2}{1-x_1}\right)\ell n\left(x_1^2\right)
       + \left(1+x_1^2\right)\left(\frac{\ell n(1-x_1)}{1-x_1}\right)_+
       - \frac{3}{2}\left(\frac{1}{1-x_1}\right)_+ \nonumber \\
&&\quad\quad + \delta(1-x_1)\left(\frac{2\pi^2}{3}-\frac{9}{2}\right)
       - \frac{1}{2}\left(3x_1-5\right) \Bigg] \nonumber \\
&&+ \left(1-3\cos^2\theta_\gamma\right) \Bigg\} .
\label{ffff}
\end{eqnarray}

For the common overall normalization function $F^{PC}_q (s)$, we use an
expression that includes $\gamma,\ Z^\circ$ interference:
\begin{eqnarray}
\frac{2}{s} F^{PC}_{q}(s)=\frac{1}{s^{2}} &\Bigg[
& e_{q}^{2}
+ \left(|v_e|^2 + |a_e|^2\right) \left(|v_q|^2 + |a_q|^2\right)
  \frac{s^2}{\left( s-M^2_Z\right)^2 + M^2_Z \Gamma^2_Z} \nonumber \\
&+&  2 e_{q}v_{e}v_{q}
  \frac{s \left( s-M_{Z}^{2} \right)}
       {\left( s-M^2_Z\right)^2 + M^2_Z \Gamma^2_Z}\, \Bigg] \ .
\label{rrr}
\end{eqnarray}
The vector $(v)$ and axial-vector $(a)$ couplings are provided
in Table~\ref{table1} and Table~\ref{table2}.
We set $M_Z = $ 91.187 GeV and $\Gamma_Z =$ 2.491 GeV.
These and other constants used here are taken from Ref.~\cite{XY}.
The weak mixing angle
$\sin^2 \theta_w =$ 0.2319. For the electromagnetic
coupling strength $\alpha_{em}$, we use the solution of the first order
QED renormalization group equation
\begin{equation}
\alpha_{em}(\mu^2) =\frac{\alpha_{em}(\mu^{2}_0)}{1+
\frac{\beta_0}{4\pi}
\alpha_{em}(\mu^{2}_0) \ell n (\mu^2/ \mu^{2}_0)}\ .
\label{sss}
\end{equation}
Here $\beta_0$ is the first order QED beta function,
\begin{equation}
\beta_0=-\frac{4}{3}\sum_{f} N_c^f e_f^2\ ,
\label{beta1}
\end{equation}
with $N_c^f$ the number of colors for flavor $f$ and $e_f$ the
fractional charge of the fermions.  The sum over $f$
extends over all fermions (leptons and quarks) with mass $m_f^2<\mu^2$.
For the energy region of interest here, we do not include the top
quark in the sum in Eq.~(\ref{beta1}), and we obtain $\beta_0=-80/9$.
To fix the boundary condition in Eq.~(\ref{sss}), we let
$\alpha_{em}(M^2_Z) = 1/128$ and set $\mu_0= M_Z$.

In the $O\left( \alpha_s\right)$ contributions, Eqs.~(\ref{dddd})
and (\ref{ffff}), we employ a two-loop expression for
$\alpha_s (\mu^2)$ with quark threshold effects handled properly.
We set $\Lambda^{(4)}_{QCD} =$ 0.231 GeV.  At $\sqrt{s} = M_Z$,
this expression provides $\alpha_s \left( M^2_Z\right) = 0.112.$

At $\sqrt{s} =$ 10 GeV, the sums in Eqs.~(\ref{bbbb}), (\ref{dddd}), and
(\ref{ffff}) run over 4 flavors of quarks $(u, d, c, s)$, all assumed
massless.  At this energy, we do not include
a $b$ quark contribution in our calculation.  For $\sqrt{s} =$ 58 GeV and
91 GeV, we use 5 flavors, again assuming all massless in the short-distance
hard scattering cross sections.  At these higher energies, non-zero mass
effects for the $c$ and $b$ quarks are accommodated by
our scale choice in the fragmentation functions, discussed below.

The quark-to-photon fragmentation function that appears in Eq.~(\ref{aaaa})
and (\ref{ffff}) is expressed as
\begin{eqnarray}
z\, D_{q \rightarrow \gamma} (z,\mu_{F}^2) &=&
 \frac{\alpha _{em}(\mu^2_F)}{2\pi} \left[
  e_{q}^{2}\
  \frac{2.21-1.28z+1.29z^{2}}{1-1.63\,\ell n\left(1-z\right)}\,
  z^{0.049} +0.002 \left( 1-z \right) ^{2} z^{-1.54}
 \right] \nonumber \\
&&\times \ell n \left( \mu_{F}^{2} / \mu^{2}_0 \right).
\label{uuu}
\end{eqnarray}
The gluon-to-photon fragmentation function in Eq.~(\ref{cccc}) is
\begin{equation}
z\, D_{g \rightarrow \gamma} (z,\mu^2_{F})
= \frac{\alpha _{em}(\mu^2_F)}{2\pi}\,
  0.0243 \left( 1-z \right)
  z^{-0.97}\, \ell n \left( \mu_{F}^{2} / \mu^{2}_0 \right).
\label{vvv}
\end{equation}
These expressions for $D_{q\rightarrow \gamma}$ and
$D_{g\rightarrow \gamma}$, taken from Ref.~\cite{JFO}, are used as
a guideline for our estimates.  The physical significance of scale
$\mu_0$ is that the fragmentation function vanishes for energies less
than $\mu_0$.  For $g$ and for the $u, d, s$, and $c$ quarks, we set
$\mu_0 = \Lambda^{(4)}_{QCD}$, as in Ref.~\cite{JFO}.
For the $b$ quark we again use Eq.~(\ref{uuu}), but
we replace $\mu_0$ by the mass of the quark,
$m_b =$ 5 GeV; $D_{b\rightarrow \gamma}(z,\mu^2_F) = 0$
for $\mu_F < m_b$. We set
the fragmentation scale $\mu_F$ equal to the renormalization scale
$\mu$ for our inclusive cross sections.  In the results presented below,
we vary $\mu$ to examine the sensitivity of the cross section
to its choice.

In presenting results, we divide our inclusive cross sections by an energy
dependent cross section $\sigma_0$ that specifies the leading order total
hadronic event rate at each value of~$\sqrt{s}$:
\begin{equation}
\sigma_0 = {{4\pi s} \over {3}} \sum_q
\left[ {2\over s} F^{PC}_q (s)\ \alpha^2_{em}(s) N_c\right].
\label{www}
\end{equation}
By doing so, we can observe what fraction of the total hadronic rate is
represented by inclusive prompt photon production.

In several figures to follow, we show the predicted behavior of the
inclusive yield as a function of $E_\gamma$ and $\theta_\gamma$,
as well as the breakdown of the total yield into contributions
from various components.

In Fig.~\ref{fig11},
we present the inclusive yield as a function of $E_\gamma$ at
$\sqrt{s} =$ 91 GeV for two values of the scattering angle $\theta_\gamma$,
45$^\circ$ and 90$^\circ$.  The same results are displayed in
Fig.~\ref{fig12} as a
function of scattering angle $\theta_\gamma$ for two choices of $E_\gamma$.
In both Figs.~\ref{fig11} and \ref{fig12},
we set renormalization/fragmentation scale
$\mu = E_\gamma$. Dependence of the cross sections on $\mu$ is examined in
Fig.~\ref{fig13} at fixed $E_\gamma$.  The patterns evident in
Figs.~\ref{fig11}--\ref{fig13} at
$\sqrt{s} =$ 91 GeV are repeated with subtle differences in
Figs.~\ref{fig14}--\ref{fig16} at
$\sqrt{s} =$ 58 GeV, appropriate for experiments at TRISTAN, and in
Figs.~\ref{fig17}--\ref{fig19}
at $\sqrt{s} =$ 10 GeV, applicable for studies at CESR/CLEO.  In
Fig.~\ref{fig20}
we compare predictions at the three energies by showing the cross
section $\sigma_0^{-1}d\sigma/dx_\gamma d\Omega_\gamma$
as a function of the scaling variable $x_\gamma = 2E_\gamma/\sqrt{s}$.

Evident in Figs.~\ref{fig11}--\ref{fig19}
is the dominance of the lowest-order contribution to the
inclusive yield, Eq.~(\ref{aaaa}), at all values of $\sqrt{s}$, except at
small values of $E_\gamma/\sqrt{s}$ or at small values
of $\mu$ where the $O\left( \alpha_{em}\right)$ ``direct" contribution,
Eq.~(\ref{bbbb}), becomes larger.  Following the lowest-order contribution
in importance at modest values of $E_\gamma/\sqrt{s}$ or of $\mu$ is the
$O\left( \alpha_{em}\right)$ direct contribution.
The direct contribution falls away more rapidly with increasing $E_\gamma$
or $\mu$ than the
$O\left( \alpha_s\right)$ quark-to-photon fragmentation term,
Eq.~(\ref{ffff}).
Therefore, at large values of $E_\gamma/\sqrt{s}$ or $\mu$, it is the
$O\left( \alpha_s\right)$ fragmentation term, that is secondary in
importance to the lowest-order term.  The gluon-to-photon fragmentation
contribution, Eq.~(\ref{cccc}) plays an insignificant role except
at very small $E_\gamma$.

In Figs.~\ref{fig12}, \ref{fig15}, and \ref{fig18},
we examine the predicted $\theta_\gamma$
dependence of our cross sections.  These figures, presented with a linear
scale, show perhaps more clearly the importance of the roles of the
$O\left( \alpha_{em}\right)$ direct and $O\left( \alpha_s\right)$
fragmentation contributions.  The lowest-order contribution,
Eq.~(\ref{aaaa}), is proportional to $(1 + \cos^2 \theta_\gamma)$.
However, there are significant $\sin^2 \theta_\gamma$ components
in the next-to-leading order direct term,
Eq.~(\ref{bbbb}), and the next-to-leading order fragmentation terms,
Eqs.~(\ref{dddd}) and (\ref{ffff}).
The net result is that the predicted total yield in
Figs.~\ref{fig12}, \ref{fig15}, and \ref{fig18}
is \underline{not} proportional to $(1 + \cos^2 \theta_\gamma)$.
As illustrated in the figures, the deviation of
the total yield from the $(1 + \cos^2 \theta_\gamma)$ form becomes
greater at smaller values of $E_\gamma$.  (The results shown in
Figs.~\ref{fig12}, \ref{fig15}, and \ref{fig18}
all pertain to the scale choice $\mu = E_\gamma$.)  One lesson from this
examination of dependence of $\theta_\gamma$ is that it is inappropriate
and potentially misleading to assume that the functional form
$(1 + \cos^2 \theta_\gamma)$ describes the data when attempts are made to
correct distributions in the region of small $\theta_\gamma$
(where initial state bremsstrahlung overwhelms the final state
radiation in which one is interested).

Dependence on the renormalization/factorization scale $\mu$ in
Figs.~\ref{fig13}, \ref{fig16}, and \ref{fig19}
shows several interesting features.  As is expected from the
functional form of $D_{q\rightarrow \gamma} \left( z, \mu^2\right)$ in
Eq.~(\ref{uuu}), the lowest-order contribution, Eq.~(\ref{aaaa}), increases
logarithmically as  $\mu$ is increased.  On the other hand, the
$\ell n \left(s/\mu^2\right)$ dependent term in Eq.~(\ref{bbbb})
causes a decrease of the
$O\left( \alpha_{em}\right)$ direct contribution as $\mu$ is increased.
Indeed, the $\left( 1 + \cos^2 \theta_\gamma\right)$ part of the direct
contribution becomes negative when
$s x^2_\gamma\left( 1-x_\gamma\right)/\mu^2
< 1$.  The physical cross  section, represented as a solid line in
Figs.~\ref{fig13}, \ref{fig16}, and \ref{fig19},
is of  course always positive.

An especially noteworthy feature of
Figs.~\ref{fig13}, \ref{fig16}, and \ref{fig19} is that the
total inclusive yield is nearly independent of $\mu$, in spite of the
strong variation with $\mu$ of its components.
This independence reflects the role of
the fragmentation scale $\mu$.  It is introduced to separate ``soft" and
``hard" contributions into ``fragmentation" and ``direct" pieces.  As the
scale $\mu$ is increased, more of the cross section is necessarily
factored into the fragmentation contribution, and vice versa,
such that the sum remains nearly constant.

In Fig.~\ref{fig20},
we show the overall $\sqrt{s}$ dependence of our predictions.  To
facilitate comparison, we present these results in terms of the ``scaling"
distribution $\sigma_0^{-1} d\sigma/dx_\gamma d\Omega_\gamma$.
The case of $\sqrt{s}=10$~GeV is somewhat special since we do not
include a contribution from b quark fragmentation at this energy.
Otherwise, the contribution of the lowest-order process,
Eq.~(\ref{aaaa}), decreases at fixed $x_\gamma$ as $\sqrt{s}$ is
increased.  This decrease is explained easily.  In computing
$d\sigma/dx_\gamma d\Omega_\gamma$, we multiply Eqs.~(\ref{rrr}) and
(\ref{uuu}), obtaining a charge weighting factor of $e_q^2 F_q^{PC}(s)$,
whereas in computing the denominator $\sigma_0$, the factor is
$F_q^{PC}(s)$.  Owing to the values of the $v$ and $a$ couplings in
Table~1, the up-type quark contribution to $F_q^{PC}(s)$ decreases as
$\sqrt{s}$ increases, and the down-type contribution increases.  The
$O(\alpha_{em})$ direct contribution to $\sigma_0^{-1} d\sigma/dx_\gamma
d\Omega_\gamma$ decreases at fixed $x_\gamma$ as $\sqrt{s}$ is increased
from 10 to 91 GeV.  Again, the explanation may be found in the energy
dependence of the ratio $\sum_q e_q^2 F_q^{PC}(s)/\sum_q F_q^{PC}(s)$.
Taken together these statements explain the energy dependence displayed
in Fig.~\ref{fig20}.

As remarked earlier, the particular expressions we chose for the
fragmentation functions are not meant to be anything but illustrative
expressions.  It would be very valuable if these non-perturbative
functions could be determined directly from data.
Dominance of the $q\rightarrow \gamma$ fragmentation contribution in
Figs.~\ref{fig11}--\ref{fig19}
demonstrates the important role data from
$e^+e^- \rightarrow \gamma X$ may play in the extraction of
$D_{q\rightarrow \gamma}\left( z, \mu^2\right)$ and study of its properties.
However, as mentioned in the Introduction, an important limitation of high
energy investigations is that photons are observed and cross sections are
measured reliably only when the photons are relatively isolated.  Since
fragmentation is a process in which photons are part of quark, antiquark,
and gluon jets, isolation reduces the contribution from fragmentation terms.
In a forthcoming paper \cite{BXQ2},
we will examine in detail the behavior of the \underline{isolated}
prompt photon cross section.

In this paper we have presented a unified treatment of inclusive prompt
photon production in hadronic final states in $e^+e^-$ annihilation.
We have computed analytically the direct photon contribution through
$O\left( \alpha_{em}\right)$ and the quark-to-photon and gluon-to-photon
fragmentation terms through $O\left( \alpha_s\right)$.  We presented the
full angular dependence of the cross section, separated into transverse
$\left( 1 + \cos^2 \theta_\gamma \right)$ and longitudinal components.

\section*{Acknowledgements}

X. Guo and J. Qiu are grateful for the hospitality of Argonne National
Laboratory where a part of this work was completed.  Work in the High
Energy Physics Division at Argonne National Laboratory is supported by
the U.S. Department of Energy, Division of High Energy Physics,
Contract W-31-109-ENG-38.
The work at Iowa State University was supported in part
by the U.S. Department of Energy under Grant Nos.
DE-FG02-87ER40731 and DE-FG02-92ER40730.

\appendix
\section{Two and Three Particle Phase Space}

In this Appendix, we express two- and three-particle final
state phase space $dPS^{(2)}$ and $dPS^{(3)}$ in
$n$ dimensions in terms of the variables necessary for our
calculation.  We work out first the specific case
of $e^+e^- \rightarrow q\bar{q}$.  The four-vector momenta of
$q$ and $\bar{q}$ are $p_q$ and $p_{\bar{q}}$.

The two particle phase space element in $n=4-2\epsilon$ dimensions is
\begin{equation}
dPS^{(2)} = {{d^{n-1}p_q} \over {(2\pi)^{n-1}2 E_q}} \cdot
            {{d^{n-1}p_{\bar{q}}} \over {(2\pi)^{n-1}2 E_{\bar{q}}}}
            \cdot (2\pi)^n \delta^{(n)} (q-p_q-p_{\bar{q}}).
\label{A.1}
\end{equation}
In the center of mass frame, $\vec{p}_q = -\vec{p}_{\bar{q}}$ and
$E_q = E_{\bar{q}}$.  Eliminating the $d^{n-1} p_{\bar{q}}$
integration, we obtain
\begin{equation}
dPS^{(2)} = \left( {{1} \over {2\pi}}\right)^{n-2} {{1} \over {8}}\,
 E^{n-4}_q dE_q\ d\theta\, \sin^{n-3}\theta d\Omega_{n-3}(p_q)\,
 \delta\left( E_q - {1\over 2}\sqrt{s} \right).
\label{A.2}
\end{equation}
Since the square of the invariant matrix element, Eq.~(\ref{y})
of the text, depends on $\theta$
but not on other angles, we may perform the integration over
$d\Omega_{n-3}$;
\begin{equation}
\Omega_{n-3} = 2\pi\ \pi^{-\epsilon}/\Gamma (1-\epsilon).
\label{A.3}
\end{equation}
We derive
\begin{equation}
dPS^{(2)} = {{1} \over {2}}\ {{1} \over {(2\pi)^3}} {{d^3p_q} \over {E_q}}
  \left[ \left( {{4\pi} \over {(s/4)\sin^2\theta}}\right)^\epsilon
         {{1} \over {\Gamma{(1-\epsilon)}}} \right]\
  {{2\pi} \over {s}}\ {{\delta(x_q -1)} \over {x_q}} ,
\label{A.4}
\end{equation}
with $x_q = 2E_q/\sqrt{s}$.

For the three particle final state
$e^+e^- \rightarrow q\bar{q}\gamma$,  we label the four-vector momenta of
$q, \bar{q},$ and $\gamma$ as $p_1, p_2$, and~$\ell$.
The invariant matrix element of interest to us,
as defined in Eqs.~(\ref{qq}) and (\ref{h2g}),
depends explicitly on the inner products
$p_1\cdot \ell, p_2\cdot \ell$, and $p_1\cdot p_2$
as well as on $p_1\cdot k$, $p_2\cdot k$, and $\ell \cdot k$ where $k$,
defined in Eq.~(\ref{r}), is the difference $k = k_{e^+}- k_{e^-}$ of the
four-momenta of the initial $e^+$ and $e^-$.  However, all these
inner products are not independent.  Using momentum conservation and the
fact that the momenta $\ell$ and $k$ are observables, one may show
that $p_1\cdot\ell$ and $p_2\cdot k$ are the only independent invariants.
Note that it is completely equivalent to choose $p_2$ instead of $p_1$.

For general orientation, it is useful to begin in $n = 4$ dimensions to
establish the\linebreak
 angular variables of integration we would use in that case,
before generalizing to $n$ dimensions.  In the overall $e^+e^-$ center
of mass frame, we imagine a coordinate system with the $\gamma$
defining the $z$ axis, vector $\vec{k}$ lying in the $(x, z)$ plane,
and vector $\vec{p}_1$ generally having non-zero $x, y$ and $z$ components.
\begin{eqnarray}
\vec{k} &=& |\vec{k}| \left(\sin\theta_\gamma, 0, \cos\theta_\gamma\right);
\label{A.5} \\
\vec{p}_1 &=& |\vec{p}_1| \left( \sin\theta_{1\gamma}\cos\phi,\,
             \sin\theta_{1\gamma} \sin\phi,
             \cos\theta_{1\gamma}\right);
\label{A.6} \\
p_1 \cdot k &=& - \vec{p}_1 \cdot \vec{k} \nonumber \\
&=& - |\vec{p}_1| |\vec{k}|
  \left( \sin\theta_\gamma\sin\theta_{1\gamma}\cos\phi
+ \cos\theta_\gamma\cos\theta_{1\gamma}\right).
\label{A.7}
\end{eqnarray}
The four-dimensional example shows that only the components of
$\vec{p}_1$ in the $\vec{\ell},\ \vec{k}$ plane contribute to
$\vec{p}\cdot \vec{k}$.
We use $\theta_x$ to denote the $n$-dimensional generalization of the
four-dimensional azimuthal angular variable $\phi$,
and we will express $dPS^{(3)}$ in $n$-dimensions in terms of
integrations over $\theta_{1\gamma}$ and $\theta_x$.

Three particle phase space in $n$-dimensions is
\begin{equation}
dPS^{(3)} =
{d^{n-1}p_1 \over (2\pi)^{n-1}2E_1}\
{d^{n-1}p_2 \over (2\pi)^{n-1}2E_2}\
{d^{n-1}\ell \over (2\pi)^{n-1}2E_\gamma}\
(2\pi)^n \delta^{(n)}(q-p_1 -p_2-\ell).
\label{A.8}
\end{equation}
Using the $\delta^{(n)}$ function to eliminate the integrations over
$p_2$, we obtain
\begin{equation}
dPS^{(3)} =
{d^{n-1}p_1 \over (2\pi)^{n-1} 2E_1}\
{d^{n-1}\ell  \over (2\pi)^{n-1}2E_\gamma}\
{1 \over 2E_2}\
2\pi\, \delta(\sqrt{s} - E_\gamma - E_1 - E_2).
\label{A.9}
\end{equation}
Since we are interested ultimately in the invariant cross section
$E_\gamma d\sigma/d^3\ell$, we rewrite Eq.~(\ref{A.9}) as follows.
\begin{eqnarray}
{d^{n-1}\ell \over (2\pi)^{n-1}2E_\gamma} &=&
{1 \over (2\pi)^{n-1}2E_\gamma}\
E_\gamma^{n-2}\ dE_\gamma\ d\theta_\gamma\ \sin^{n-3}\theta_\gamma\
d\Omega_{n-3} (\ell)\nonumber \\
&\equiv &{{1}\over{2}}\, {{1} \over {(2\pi)^3}}
{{d^3\ell} \over {E_\gamma}}\
{{1} \over {(2\pi)^{n-4}}}\
\left( E^2_\gamma \sin^2 \theta_\gamma \right)^{{n-4} \over {2}}\
{{d\Omega_{n-3}(\ell)} \over {d\phi_\gamma}}\ .
\label{A.10}
\end{eqnarray}
We take angle $\theta_\gamma$ to be the polar angle of the $\gamma$ with
respect to the $e^+e^-$ collision axis in the overall center of mass
frame.  Since the square of the matrix element does not depend on
$\phi_\gamma$, we can integrate over $d\Omega_{n-3}(\ell)$ and
$d\phi_\gamma$ independently.  Using $\Omega_{n-3}$ from
Eq.~(\ref{A.3}), and $\int d\phi_\gamma=2\pi$,
we reexpress Eq.~(\ref{A.10})
as
\begin{equation}
{d^{n-1}\ell \over (2\pi)^{n-1}2E_\gamma} =
{{1}\over{2}}\, {{1} \over {(2\pi)^3}}\, {{d^3\ell} \over {E_\gamma}}\
\left[ \left(\frac{4\pi}{E^2_\gamma\sin^2\theta_\gamma}\right)^\epsilon\,
\frac{1}{\Gamma(1-\epsilon)} \right].
\label{A.10a}
\end{equation}
We write $d^{n-1}p_1$ in Eq.~(\ref{A.9}) as
\begin{eqnarray}
d^{n-1}p_1
&=& E^{n-2}_1\ dE_1\ d\Omega_{n-2}(p_1)\nonumber \\
&=& E_1^{n-2}\ dE_1\ d\theta_{1\gamma}\
   \sin^{n-3}\theta_{1\gamma}\ d\Omega_{n-3}(p_1),
\label{A.12}
\end{eqnarray}
with
\begin{eqnarray}
d\Omega_{n-3}(p_1)
&=& d\theta_x\ \sin^{n-4}\theta_x\ d\Omega_{n-4}(p_1)\nonumber \\
&=& d\cos\theta_x\ (1-\cos^2\theta_x)^{n-5 \over 2}\
   d\Omega_{n-4}(p_1)
\label{A.13}
\end{eqnarray}
Since only the components of $p_1$ in the $\vec{\ell}$, $\vec{k}$ plane
contribute to $p_1\cdot k$, as shown in Eq.~(\ref{A.7}),
all angular variables on which the invariant matrix element
depends are displayed explicitly in Eqs.~(\ref{A.12}) and (\ref{A.13}).
We may therefore integrate $d\Omega_{n-4}(p_1)$
in Eq.~(\ref{A.13}) to obtain
\begin{equation}
\Omega_{n-4}(p_1) =
{2^{n-4}\pi^{{n-4 \over 2}}\Gamma({n-4 \over 2}) \over
\Gamma(n-4)}.
\label{A.14}
\end{equation}
In this frame, $E_2$ in Eq.~(\ref{A.9}) can be expressed as
\begin{equation}
E^2_2 = (\vec{p}_2)^2 = (\vec{p}_1 + \vec{\ell})^2 =
E^2_1 + E^2_\gamma + 2E_1 E_\gamma \cos \theta_{1\gamma}.
\label{A.11}
\end{equation}
Using Eq.~(\ref{A.11}), we can replace the integration over
$d\cos\theta_{1\gamma}$ in Eq.~(\ref{A.12})
by an integration over $dE_2$; for fixed $E_1$,
\begin{equation}
E_1 E_\gamma d\cos\theta_{1\gamma} = E_2\ dE_2\ .
\label{A.15}
\end{equation}
Substituting into Eq.~(\ref{A.9}), we derive
\begin{eqnarray}
dPS^{(3)} &=& \ {{1} \over {2}}\
 {{1} \over {(2\pi)^3}}\ {{d^3 \ell} \over {E_\gamma}}\
 \left[ \left(
        \frac{4\pi}
             {E^2_{\gamma} \sin^2 \theta_{\gamma}} \right)^\epsilon \,
 \frac{1}{\Gamma(1-\epsilon)} \right]\nonumber \\
&\times &
 {{1} \over {4}}\ {{1} \over {(2\pi)^{2}}}
 \left(\frac{4\pi^2}
            {E^2_1\sin^2\theta_{1\gamma}} \right)^{\epsilon}
 \Omega_{n-4}(p_1)\, (1-\cos^2\theta_x)^{-\epsilon-1/2}\,
 d\cos\theta_x \nonumber \\
&\times &
 {{1} \over {E_\gamma}}
 \delta(\sqrt{s}-E_\gamma-E_1-E_2)\
 dE_1\ dE_2\ .
\label{A.16}
\end{eqnarray}
It is easy to verify that Eq.~(\ref{A.16}) reduces to the familiar form in
four-dimensions when $\epsilon \rightarrow 0$.

We introduce new dimensionless variables related to the singularity
structure of the invariant matrix elements, given in
Eqs.~(\ref{qq}) and (\ref{h2g}):
\begin{mathletters}
\label{A.17}
\begin{eqnarray}
y_{12} & \equiv &{2p_1\cdot p_2 \over q^2}
\label{A.171}\\
y_{1\ell} & \equiv &{2p_1\cdot \ell \over q^2}
\label{A.172}\\
y_{2\ell} & \equiv &{2p_2\cdot \ell \over q^2}
\label{A.173}\\
x_1 & \equiv &{2p_1\cdot q \over q^2}
\label{A.174)}\\
x_2 & \equiv &{2p_2\cdot q \over q^2}
\label{A.175}\\
x_\gamma & \equiv &{2\ell \cdot q \over q^2}
\label{A.176}
\end{eqnarray}
\end{mathletters}
We observe that $y_{1\ell} = 1-x_2,\ y_{2\ell} = 1-x_1$, and
$y_{12} = 1 - x_\gamma$.  In the center of mass frame,
$q = (\sqrt{s}, \vec{0})$, we have
\begin{equation}
x_1 = {2E_1 \over \sqrt{s}}\, , \quad{\rm and}\quad
x_2 = {2E_2 \over \sqrt{s}}\ ;
\label{A.18}
\end{equation}
\begin{equation}
dE_1\ dE_2\ =\ {s \over 4}\ dx_1\ dx_2\
=\ {s \over 4}\ dy_{1\ell}\ dy_{2\ell}\ ;
\label{A.19}
\end{equation}
\begin{equation}
\delta(\sqrt{s} - E_q - E_{\bar{q}} - E_\gamma) = {2 \over \sqrt{s}}
\delta (1-y_{1\ell}-y_{2\ell} - y_{12}).
\label{A.20}
\end{equation}
After some algebra, one may verify that
\begin{equation}
E^2_1\ E^2_\gamma\ \sin^2\ \theta_{1\gamma} =\
{s^2 \over 4}\ y_{1\ell}\ y_{2\ell}\ y_{12}.
\label{A.21}
\end{equation}
Substituting Eqs. (\ref{A.19})-(\ref{A.21}) into Eq.~(\ref{A.16}), we
derive
\begin{eqnarray}
dPS^{(3)} &=& {{1} \over {2}}\
 {{1} \over {(2\pi)^3}}\ {{d^3 \ell} \over {E_\gamma}}\
 \left[ \left(\frac{4\pi}{(s/4)\sin^2\theta_\gamma}\right)^\epsilon\,
 \frac{1}{\Gamma(1-\epsilon)} \right]\
 \frac{1}{x_\gamma} \nonumber \\
&\times &
 {{1} \over {4}}\ {{1} \over {(2\pi)^{2}}}
 \left(\frac{4\pi^2}{s}\right)^{\epsilon}
 \Omega_{n-4}(p_1)\, (1-\cos^2\theta_x)^{-\epsilon-1/2}\,
 d\cos\theta_x \nonumber \\
&\times &
 \frac{dy_{1\ell}\ dy_{2\ell}\ \delta(1-y_{1\ell}-y_{2\ell}-y_{12})}
      {\left(y_{1\ell}\ y_{2\ell}\ y_{12}\right)^{\epsilon} }\ .
\label{A.22}
\end{eqnarray}
For reference, we record that
\begin{equation}
\int^1_{-1}d\cos \theta_x (1-\cos^2 \theta_x)^{-\epsilon-{1 \over 2}} =
{\Gamma ({1 \over 2})\Gamma ({1 \over 2} - \epsilon) \over \Gamma
(1-\epsilon)}
\label{A.23}
\end{equation}
\begin{equation}
\int^1_{-1} d\cos \theta_x\ \cos^2 \theta_x\
(1-\cos^2\theta_x)^{-\epsilon - {1 \over 2}} =
{\Gamma({3 \over 2}) \Gamma({1 \over 2} - \epsilon) \over \Gamma
(2-\epsilon)}.
\label{A.24}
\end{equation}
Dividing Eq.~(\ref{A.24}) by Eq.~(\ref{A.23}), we define the average
of $\cos^2\theta_x$ in $n$-dimensions as
\begin{equation}
\langle \cos^2 \theta_x\rangle = {1\over 2}\ {1 \over {1-\epsilon}}.
\label{A.25}
\end{equation}
When $\epsilon\rightarrow 0$, $\theta_x$ becomes the azimuthal angular
$\phi$, and Eq.~(\ref{A.25}) is consistent with the 4-dimensional
result
\begin{equation}
\langle \cos^2\phi \rangle \equiv
\frac{1}{2\pi} \int_0^{2\pi} d\phi \cos^2\phi
=\frac{1}{2} \ .
\label{A.25a}
\end{equation}

Equation (\ref{A.22}) is written in a form for which the photon
with momentum $\ell$ is the observed particle, with integrations
done over the momenta of other final state partons.
When considering $e^+e^- \rightarrow q \bar{q} g$ with $q$
(or $\bar{q}$) fragmenting into the observed $\gamma$, we require instead
$E_q d\sigma/d^3p_q$.  It is useful, therefore, to reexpress
Eq.~(\ref{A.22}) in a form that manifests the symmetry of
phase space among all three final state particles.

Using $y_{12} = 1-x_\gamma$, we introduce the identity
\begin{equation}
1 = dy_{12}\ \delta \left(x_\gamma - (1-y_{12})\right)\ .
\label{A.26}
\end{equation}
Inserting this identity into Eq.~(\ref{A.22}), we obtain a more
symmetric form
\newpage
\begin{eqnarray}
dPS^{(3)} &=& {{1} \over {2}}\
 {{1} \over {(2\pi)^3}}\ {{d^3 \ell} \over {E_\gamma}}\
 \left[ \left(\frac{4\pi}{(s/4)\sin^2\theta_\gamma}\right)^\epsilon\,
 \frac{1}{\Gamma(1-\epsilon)} \right]\
 \frac{2\pi}{s}\,
 \frac{\delta(x_\gamma-(1-y_{12}))}{x_\gamma} \nonumber \\
&\times &
 {{s} \over {4}}\ \left[ \left(\frac{1}{2\pi}\right)^{2}
 \left(\frac{4\pi}{s}\right)^{\epsilon}
 \frac{1}{\Gamma(1-\epsilon)}
 \frac{d\Omega_{n-3}(p_1)}{\Omega_{n-3}(p_1)} \right]\nonumber \\
&\times &
 \frac{dy_{1\ell}}{y^\epsilon_{1\ell}}\
 \frac{dy_{2\ell}}{y^\epsilon_{2\ell}}\
 \frac{dy_{12}}{y^\epsilon_{12}}\
 \delta(1-y_{1\ell} - y_{2\ell}-y_{12}) \ .
\label{A.22a}
\end{eqnarray}

We can use Eq.~(\ref{A.22a}) to derive an expression for
three-particle phase-space suitable for calculating
$E_q d\sigma/d^3p_q$ when the quark from $e^+e^-\rightarrow
q\bar{q}g$ is the fragmenting parton.  We label
$(q,\bar{q},g)$ as (1,2,3).  We
let the gluon replace the photon $\gamma$ in Eq.~(\ref{A.22a}),
and we then switch the labels of the $g$ and the observed quark.
We derive
\begin{eqnarray}
dPS^{(3)} &=& {{1} \over {2}}\
 {{1} \over {(2\pi)^3}}\ {{d^3 p_1} \over {E_1}}\
 \left[ \left(\frac{4\pi}{(s/4)\sin^2\theta_1}\right)^\epsilon\,
 \frac{1}{\Gamma(1-\epsilon)} \right]\
 \frac{2\pi}{s}\,
 \frac{\delta(x_1-(1-y_{23}))}{x_1} \nonumber \\
&\times &
 {{s} \over {4}}\ \left[ \left(\frac{1}{2\pi}\right)^{2}
 \left(\frac{4\pi}{s}\right)^{\epsilon}
 \frac{1}{\Gamma(1-\epsilon)}
 \frac{d\Omega_{n-3}(p_3)}{\Omega_{n-3}(p_3)} \right]\nonumber \\
&\times &
 \frac{dy_{12}}{y^\epsilon_{12}}\
 \frac{dy_{13}}{y^\epsilon_{13}}\
 \frac{dy_{23}}{y^\epsilon_{23}}\
 \delta(1-y_{12} - y_{13}-y_{23}) \ .
\label{A.27}
\end{eqnarray}
The first line of Eq.~(\ref{A.27}) is identical to $dPS^{(2)}$,
Eq.~(\ref{A.4}).  By switching $1$ and $2$ in Eq.~(\ref{A.27}), one can
get a form for the phase space suitable for calculating the hard-scattering
cross section when the antiquark is the fragmenting parton.



\begin{figure}
\caption{Illustration of $e^+e^- \rightarrow \gamma X$ in an $m$ parton
state: $e^+e^- \rightarrow c X$ followed by fragmentation
$c \rightarrow \gamma X$.}
\label{fig1}
\end{figure}

\begin{figure}
\caption{Contribution to the hard-scattering cross section for
$e^+e^- \rightarrow c X$ with the fragmentation of
$c \rightarrow \gamma X$.  Here $c$ denotes an intermediate photon or a
gluon or a quark of any flavor.}
\label{fig2}
\end{figure}

\begin{figure}
\caption{Lowest order, $O\left( \alpha^o_{em}\alpha^o_s\right)$, photon
production through quark fragmentation.}
\label{fig3}
\end{figure}

\begin{figure}
\caption{Feynman diagrams for $e^+e^- \rightarrow \gamma q\bar{q}$.}
\label{fig4}
\end{figure}

\begin{figure}
\caption{Order $\alpha_s$ Feynman diagrams for
$e^+e^- \rightarrow q\bar{q} g$ that contribute to
$e^+e^- \rightarrow \gamma X$ via $g\rightarrow \gamma$ fragmentation.}
\label{fig5}
\end{figure}

\begin{figure}
\caption{Contributions to the $O(\alpha_s)$ cross section
$\sigma^{(1)}_{e^+e^- \rightarrow q X}$; (a) real gluon emission diagrams
$(e^+e^- \rightarrow q\bar{q}g)$, (b) virtual gluon exchange diagrams
that interfere with the lowest order tree diagram.}
\label{fig6}
\end{figure}

\begin{figure}
\caption{Order $O(\alpha_{em}^2)$ contribution to the
hadronic tensor $H_{\mu\nu}$.}
\label{fig7}
\end{figure}

\begin{figure}
\caption{Center of mass coordinate axes of an
$e^+e^-$ collision with the $z$-axis
being the direction of the observed photon.}
\label{fig8}
\end{figure}

\begin{figure}
\caption{Feynman diagram for $D^{(1)}_{q\rightarrow\gamma}(z)$.}
\label{fig9}
\end{figure}

\begin{figure}
\caption{Feynman diagrams for $D^{(1)}_{q\rightarrow q}(z)$.}
\label{fig10}
\end{figure}

\begin{figure}
\caption{Normalized invariant cross section for the inclusive preocess
$e^+e^-\rightarrow\gamma X$ at $\protect \sqrt{s}=91$~GeV
shown as a function of
the photon energy $E_\gamma$ for two values of the photon scattering
angles: a) $\theta_\gamma =45^0$, and b) $\theta_\gamma = 90^0$.
Displayed are the total result and the four separate contributions from
lowest-order fragmentation (``oth-frag''), $O(\alpha_{em})$ direct
production, and the $O(\alpha_s)$ quark and gluon fragmentation
contributions.
The renormalization/fragmentation scale $\mu=E_\gamma$. }
\label{fig11}
\end{figure}

\begin{figure}
\caption{Normalized invariant cross section for the inclusive preocess
$e^+e^-\rightarrow\gamma X$ at $\protect \sqrt{s}=91$~GeV
shown as a function of
the photon scattering angle $\theta_\gamma$
for two values of the photon energy:
a) $E_\gamma =15$~GeV, and b) $E_\gamma = 30$~GeV.
The total result and the four separate component
pieces are displayed.
The renormalization/fragmentation scale $\mu=E_\gamma$. }
\label{fig12}
\end{figure}

\begin{figure}
\caption{Renormalization/factorization scale dependence of the
normalized invariant cross section for the inclusive process
$e^+e^-\rightarrow\gamma X$ at $\protect \sqrt{s}=91$~GeV
for two values of the photon scattering
angle: a) $\theta_\gamma =45^0$, and b) $\theta_\gamma = 90^0$.
The photon energy $E_{\gamma} = 15$~GeV.
The total result shows little $\mu$ dependence, whereas the component
contributions display considerable compensating variation with $\mu$.
}
\label{fig13}
\end{figure}

\begin{figure}
\caption{Photon energy dependence, as in Fig.~\protect\ref{fig11},
but for center-of-mass energy $\protect \sqrt{s}=58$~GeV.
}
\label{fig14}
\end{figure}

\begin{figure}
\caption{Photon scattering angle dependence, as in Fig.~\protect\ref{fig12},
but for center-of-mass energy $\protect \sqrt{s}=58$~GeV,
and photon energies
a) $E_\gamma =10$~GeV, and b) $E_\gamma = 20$~GeV.
}
\label{fig15}
\end{figure}

\begin{figure}
\caption{Renormalization/factorization scale dependence, as in
Fig.~\protect\ref{fig13}, but for center-of-mass energy
$\protect \sqrt{s}=58$~GeV,
and photon energy $E_\gamma = 10$~GeV.
}
\label{fig16}
\end{figure}

\begin{figure}
\caption{Photon energy dependence, as in Figs.~\protect\ref{fig11}
and \protect\ref{fig14}, but for
center-of-mass energy $\protect\sqrt{s}=10$~GeV.
}
\label{fig17}
\end{figure}

\begin{figure}
\caption{Photon scattering angle dependence, as in Figs.~\protect\ref{fig12}
and \protect\ref{fig15}, but for center-of-mass energy
$\protect\sqrt{s}=10$~GeV, and photon energies
a) $E_\gamma =1.5$~GeV, and b) $E_\gamma = 3$~GeV.
}
\label{fig18}
\end{figure}

\begin{figure}
\caption{Renormalization/factorization scale dependence, as in
Figs.~\protect\ref{fig13} and \protect\ref{fig16},
but for center-of-mass energy
$\protect\sqrt{s}=10$~GeV,
and photon energy $E_\gamma = 2$~GeV.
}
\label{fig19}
\end{figure}

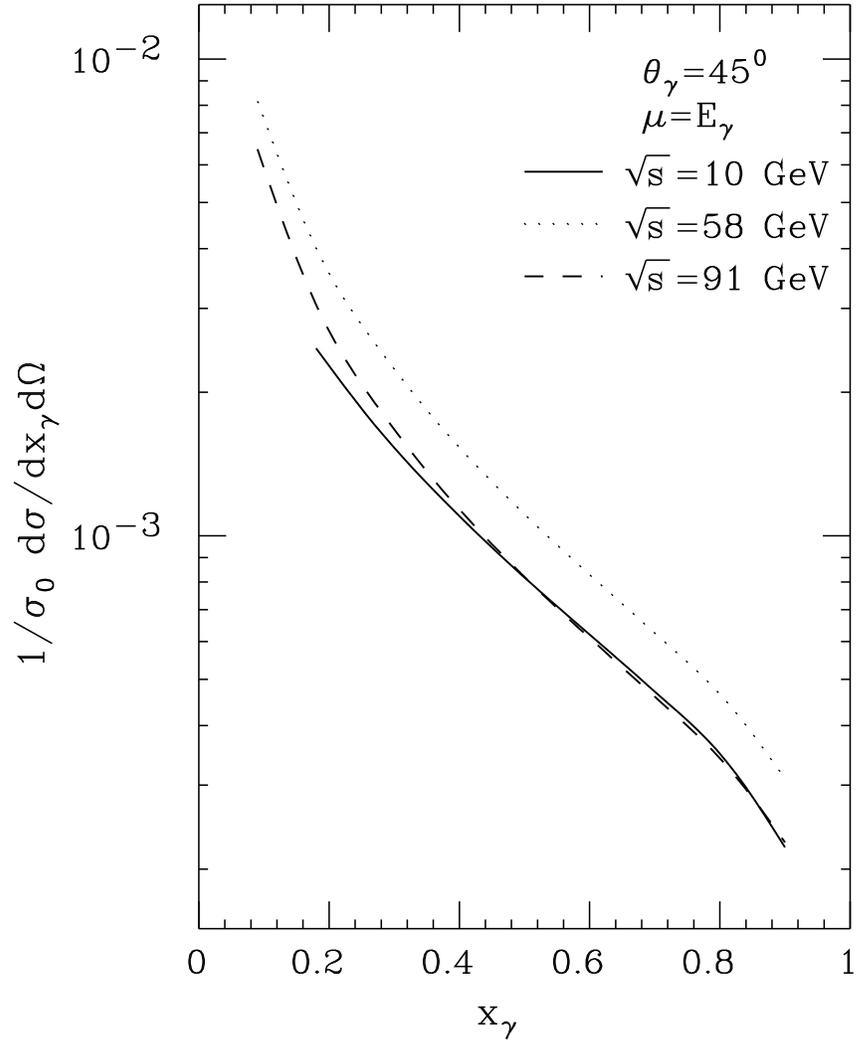
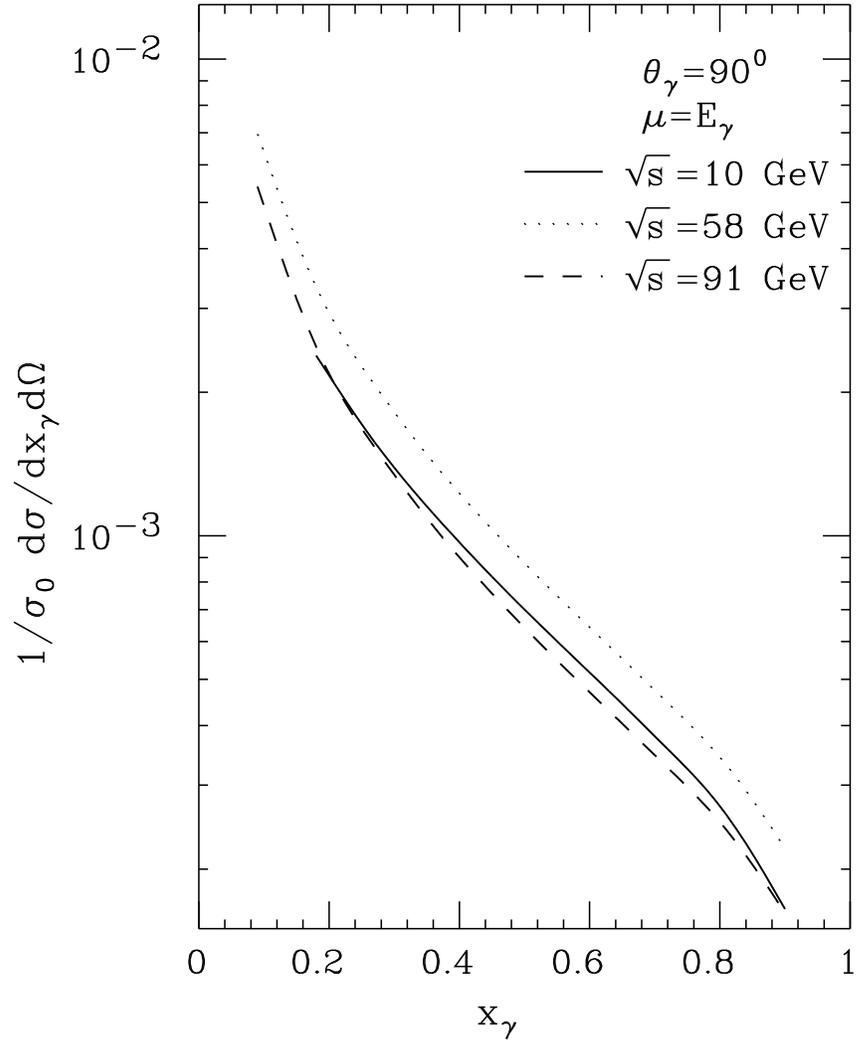
\begin{figure}
\caption{Normalized invariant cross section for the inclusive preocess
$e^+e^-\rightarrow\gamma X$ expressed in terms of the scaling variable
$x_\gamma=2E_\gamma/\protect\sqrt{s}$.
Results are presented for two values of
the photon scattering angle:
a) $\theta_\gamma =45^0$, and b) $\theta_\gamma = 90^0$.  Shown are
curves for three center-of-mass energies:
$\protect\sqrt{s}=$10, 58, and 91~GeV.
}
\label{fig20}
\end{figure}


\narrowtext
\begin{table}
\caption{Electroweak V-A coupling constants}
\vspace{0.2in}
\begin{tabular}{cc}
$v_{e}$ & $\left(-1+4 \sin^{2}\theta_{w}\right)/
           \left(2\sin 2\theta_{w}\right)$
        \\ \hline
$a_{e}$ & $1/\left(2\sin 2\theta_{w}\right)$
        \\ \hline
$v_{q}$ & $\left(I_{3}^{q}-2e_{q}\sin^{2}\theta_{w}\right)/
           \left(\sin 2\theta_{w}\right)$
        \\ \hline
$a_{q}$ & $-I_{3}^{q}/\sin 2\theta_{w}$ \\
\end{tabular}
\label{table1}
\end{table}

\begin{table}
\caption{Isospin and fractional charges for quarks}
\vspace{0.2in}
\begin{tabular}{ccc}
       & $I_{3}^{q}$    & $e_{q}$        \\ \hline
u,\ c,\ t  & $ 1/2 $ & $ 2/3 $  \\ \hline
d,\ s,\ b  & $-1/2 $ & $-1/3 $ \\
\end{tabular}
\label{table2}
\end{table}

\end{document}